# Rate Regions for Relay Broadcast Channels [1] [2]

Yingbin Liang and Gerhard Kramer [3]

**Abstract**

A partially cooperative relay broadcast channel (RBC) is a three-node network with one source node and two destination nodes (destinations 1 and 2) where destination 1 can act as a relay to assist destination 2. Inner and outer bounds on the capacity region of the discrete memoryless partially cooperative RBC are obtained. When the relay function is disabled, the inner bound reduces to an inner bound on the capacity region of broadcast channels that is at least as large as Marton's inner bound. The outer bound reduces to a new outer bound on the capacity region of broadcast channels that generalizes an outer bound of Marton to include a common message, and that generalizes an outer bound of Gel'fand and Pinsker to general discrete memoryless broadcast channels. Our proof for this outer bound simplifies the proof of Gel'fand and Pinsker that was based on a recursive approach. Four classes of RBCs are studied in detail. For the *partially cooperative RBC with degraded message sets*, inner and outer bounds are obtained. For the *semideterministic partially cooperative RBC* and the *orthogonal partially cooperative RBC*, the capacity regions are established. For the *parallel partially cooperative RBC with unmatched degraded subchannels*, the capacity region is established for the case of degraded message sets. The capacity is also established when the source node has only a private message for destination 2, i.e., the channel reduces to a *parallel relay channel with unmatched degraded subchannels*.

## 1 Introduction

Relay broadcast channels (RBCs) are communication networks where a source node transmits information to a number of destination nodes that cooperate by exchanging information. These channels model "downlink" communication systems that incorporate relaying and user cooperation to achieve higher throughput. Three RBC models have been studied recently. In [1], a two-destination partially cooperative RBC model (see Fig. 1 (a)) was studied, where one destination node (destination 1) acts as a relay [2, 3] and transmits cooperative information to the other destination node (destination 2). It was shown that the capacity region of the original broadcast channel (BC) was improved due to relaying and user cooperation. This RBC model was further studied in [4] for the case of more than two destinations, where bounds on the capacity region and minimum energy per bit were derived. The fully cooperative RBC (see Fig. 1 (b)) is a more general model studied in [1] where both destinations relay information to each other. The partially and fully cooperative RBCs were also studied in [5] for the case where the relay-to-relay (or destination-to-destination) channels are orthogonal to each other and to the broadcast channel. A third RBC model called the dedicated-relay BC (see Fig. 1 (c)) was studied in [6] and [1], where an additional relay node was introduced to the broadcast channel to assist all destination nodes.

---

[1]The material in this paper was presented in part at the Annual Conference on Information Sciences and Systems (CISS), Princeton, New Jersey, March 2006.

[2]The work of Y. Liang was supported by a Vodafone Foundation Graduate Fellowship. The work of G. Kramer was partially supported by the Board of Trustees of the University of Illinois Subaward No. 04-217 under NSF Grant No. CCR-0325673.

[3]Yingbin Liang was with the Department of Electrical and Computer Engineering and the Coordinated Science Laboratory, University of Illinois at Urbana-Champaign, Urbana IL 61801. She is now with the Department of Electrical Engineering, Princeton University, Engineering Quadrangle, Olden Street, Princeton, NJ 08544; e-mail: `yingbinl@princeton.edu`; Gerhard Kramer is with the Bell Laboratories, Lucent Technologies, Murray Hill, NJ 07974; e-mail: `gkr@research.bell-labs.com`

For the partially cooperative RBC, the achievable rate regions given in [1] are based on the source node using superposition encoding and the relay employing a decode-and-forward scheme. These encoding schemes were shown to be optimal for certain cases, e.g., for degraded partially cooperative RBCs, physically degraded Gaussian channels, and feedback channels. The achievable regions given in [4] are based on the source node using either superposition encoding or binning and the relay using a decode-and-forward scheme. In general, these encoding schemes are not optimal.

In this paper, we derive improved inner and outer bounds on the capacity region of the partially cooperative RBC. For the inner bounds, the source node uses superposition encoding and binning, and the relay employs a partial decode-and-forward scheme. Our outer bound is different from the outer bounds given in [1] and is based on a different approach. By choosing the relay channel inputs to be null, both our inner and outer bounds reduce to new bounds on the capacity region of broadcast channels. In particular, our new inner bound on the broadcast channel capacity region is at least as large as Marton's inner bound [7]. Our new outer bound generalizes Marton's outer bound [7] to include common messages, and generalizes Gel'fand and Pinsker's outer bound [8] to general discrete memoryless broadcast channels. Furthermore, our proof for the bound is different from the proof given in [8] that was based on a recursive approach.

We study four classes of partially cooperative RBCs in detail. The first channel is the *partially cooperative RBC with degraded message sets*, where the source node has a common message for both destinations and a private message for destination 1. For this channel, our inner and outer bounds have the same form. However, the joint distributions for the two bounds satisfy different Markov chain conditions, and thus prevent us from claiming we have found the capacity region.

The second channel is the *semideterministic partially cooperative RBC*, where the relay (destination 1) output is a deterministic function of the source and relay inputs. This model is a generalization of the semideterministic broadcast and relay channels studied in [8] and [9], respectively. We establish the capacity region for this channel and illustrate our result via an example channel, which we call the Blackwell partially cooperative RBC.

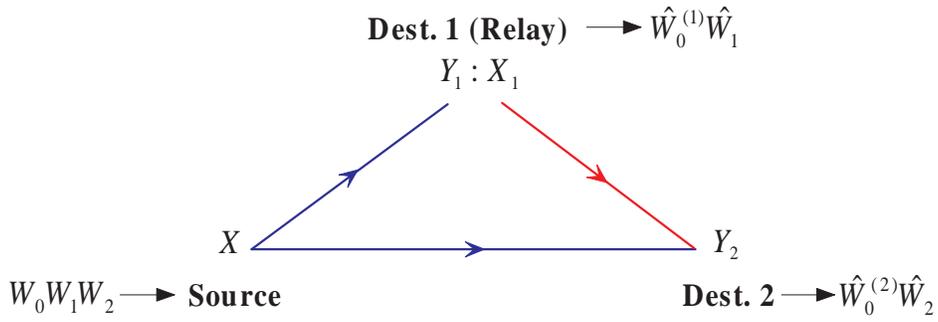

(a) Partially Cooperative RBC (see also Fig. 2)

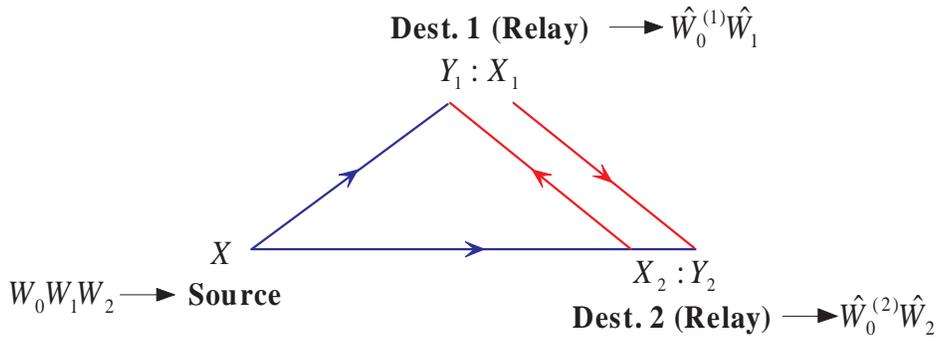

(b) Fully Cooperative RBC

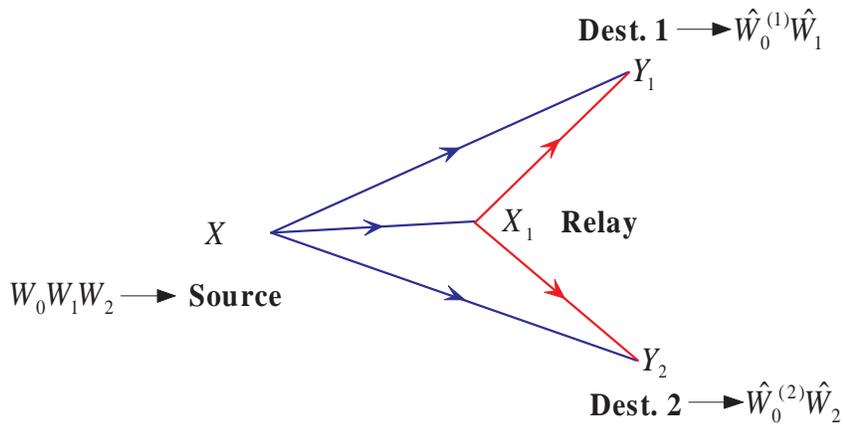

(c) Dedicated-relay BC

Figure 1: Three Relay Broadcast Channels

The third channel we study is the *orthogonal partially cooperative RBC*, where the source transmits to destination 1 in one channel and the source and relay transmit to destination 2 in another orthogonal channel. This model is a generalization of the relay channel with orthogonal components studied in [10]. For this channel, we establish the capacity region for both the discrete memoryless and Gaussian cases.

The fourth channel we study is the *parallel partially cooperative RBC with unmatched degraded subchannels*, which consists of two degraded partially cooperative RBCs as subchannels from the source node to the two destination nodes. In the first subchannel, the output of destination 2 is a degraded version of the output of destination 1, and in the second subchannel, the output of destination 1 is a degraded version of the output of destination 2. Our inner and outer bounds on the capacity region do not match in general for this channel. Three cases are further studied. For the case of degraded message sets, i.e., where the source has a common message for both destinations and a private message for destination 1, we establish the capacity region. For case 2 where the source has only a private message for destination 2, i.e., the channel reduces to the *parallel relay channel with unmatched degraded subchannels*, we obtain the capacity. For case 3 where the source has only private messages for the destinations, we show that the achievable region is larger than the Minkowski sum of the capacity regions of the two subchannels. This is in contrast to the capacity of parallel broadcast channels with unmatched degraded subchannels [11, 12].

The organization of this paper is as follows. In Section 2, we describe the partially cooperative RBC model. In Section 3, we present inner and outer bounds on the capacity region of the partially cooperative RBC that give new bounds on the capacity region of the broadcast channel. In Sections 4-7, we present our results for four classes of channels, namely the partially cooperative RBC with degraded message sets, the semideterministic partially cooperative RBC, the orthogonal partially cooperative RBC, and the parallel partially cooperative RBC with unmatched degraded subchannels. Section 8 concludes the paper. Note that Sections 3-6 appeared in the Ph.D. thesis of the first author [13, Chap. 5].

## 2   Channel Model

The partially cooperative RBC consists of a source input alphabet $\mathcal{X}$, a relay input alphabet $\mathcal{X}_1$, and two channel output alphabets $\mathcal{Y}_1$ and $\mathcal{Y}_2$. The channel is characterized by the probability distribution $p(y_1, y_2|x, x_1)$, where $x$ indicates the source input, $x_1$ indicates the relay input of destination 1, and $y_1$ and $y_2$ indicate the outputs at destinations 1 and 2, respectively. We assume that the channel is memoryless, i.e., the present channel outputs depend on the messages, the previous channel inputs, and the previous channel outputs only through the present channel inputs.

Fig. 2 illustrates the channel model. The source has a common message $W_0$ that is decoded by both destinations, and private messages $W_1$ and $W_2$ that are decoded by destinations 1 and 2, respectively. The channel reduces to a broadcast channel [14] if destination 1 (the relay) does not relay information to destination 2. The channel reduces to the relay channel [2, 3] if the source has only a private message $W_2$ intended for destination 2. As in Fig. 2, we use upper case letters to indicate random variables and lower case letters to indicate realizations of random variables. Exceptions will be clarified where they appear in the paper. We use $x^n$ and $x_i^n$ to represent the vectors $(x_1, \ldots, x_n)$ and $(x_i, \ldots, x_n)$, respectively.

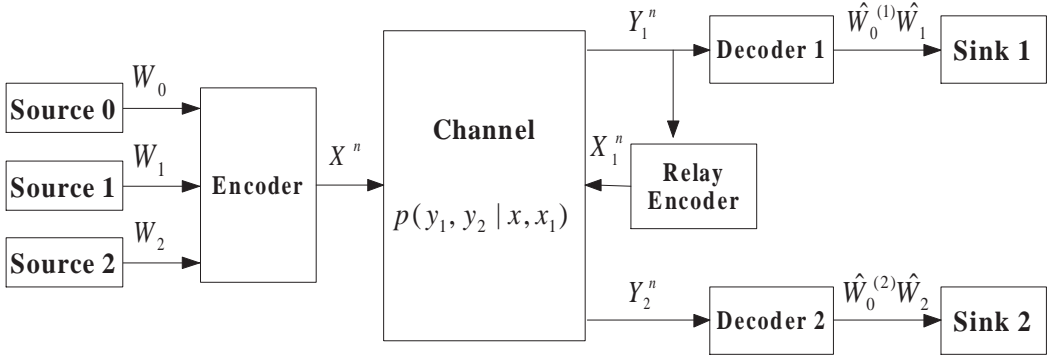

Figure 2: Partially cooperative RBC.

A communication strategy for $n$ channel uses and with rate-triple $(R_0, R_1, R_2)$ consists of the following:

- Three message sets $\mathcal{W}_k = \{1, 2, \ldots, 2^{nR_k}\}$, $k = 0, 1, 2$;

- Three independent messages $W_0$, $W_1$ and $W_2$ that are uniformly distributed over $\mathcal{W}_0$, $\mathcal{W}_1$ and $\mathcal{W}_2$, respectively;

- An encoder that maps each message triple $(w_0, w_1, w_2)$ to a codeword $x^n \in \mathcal{X}^n$;

- Deterministic relay functions $f_1, f_2, \ldots, f_n$ such that $x_{1i} = f_i(y_{11}, \ldots, y_{1[i-1]})$ for $1 \leq i \leq n$;

- Two decoders: destination $k$, $k = 1, 2$, maps $y_k^n$ to a pair $(\hat{w}_0^{(k)}, \hat{w}_k) \in \mathcal{W}_0 \times \mathcal{W}_k$.

The probability of error when the message triple $(W_0, W_1, W_2)$ is sent is defined as

$$P_e(W_0, W_1, W_2) = Pr\left\{(\hat{W}_0^{(1)}, \hat{W}_1) \neq (W_0, W_1) \text{ or } (\hat{W}_0^{(2)}, \hat{W}_2) \neq (W_0, W_2)\right\} \quad (1)$$

and the average block probability of error is

$$P_e = \frac{1}{2^{nR_0} 2^{nR_1} 2^{nR_2}} \sum_{W_0=1}^{2^{nR_0}} \sum_{W_1=1}^{2^{nR_1}} \sum_{W_2=1}^{2^{nR_2}} P_e(W_0, W_1, W_2). \quad (2)$$

The rate triple $(R_0, R_1, R_2)$ is said to be *achievable* for the partially cooperative RBC if, for any $\epsilon > 0$, there is a (sufficiently long) code with $P_e \leq \epsilon$. The *capacity region* is the closure of the set of all achievable rate triples.

# 3 Bounds for Partially Cooperative RBCs and Broadcast Channels

We use superposition encoding [15] and binning [16] at the source node, and the decode-and-forward scheme [3] for the common message $W_0$. The result is Theorem 1. We then enlarge our achievable region in Theorems 2 and 3. The resulting regions are using the decode-and-forward scheme for the common message $W_0$ and a partial decode-and-forward scheme [3, 9] for the private message $W_2$.

**Theorem 1.** *The following is an achievable region for the partially cooperative RBC:*

$$\mathcal{R}'_1 = \bigcup_{\substack{p(x_1, t, u_1, u_2, x) \\ p(y_1, y_2 | x_1, x)}} \left\{ \begin{array}{l} (R_0, R_1, R_2): \\ R_0 \geq 0, R_1 \geq 0, R_2 \geq 0, \\ R'_1 \geq 0, R'_2 \geq 0, \\ R_1 + R'_1 < I(U_1; Y_1 | T, X_1), \\ R_0 + R_1 + R'_1 < I(T, U_1; Y_1 | X_1), \\ R_2 + R'_2 < I(U_2; Y_2 | T, X_1), \\ R_0 + R_2 + R'_2 < I(T, X_1, U_2; Y_2), \\ R'_1 + R'_2 > I(U_1; U_2 | T, X_1), \end{array} \right\}. \quad (3)$$

*Furthermore, the region $\mathcal{R}'_1$ is equivalent to the following region:*

$$\mathcal{R}_1 = \bigcup_{p(x_1, t, u_1, u_2, x) p(y_1, y_2 | x_1, x)}$$
$$\left\{ \begin{array}{l} (R_0, R_1, R_2): \\ R_0 \geq 0, R_1 \geq 0, R_2 \geq 0, \\ R_1 < I(U_1; Y_1 | T, X_1), \\ R_0 + R_1 < I(T, U_1; Y_1 | X_1), \\ R_2 < I(U_2; Y_2 | T, X_1), \\ R_0 + R_2 < I(T, X_1, U_2; Y_2), \\ R_1 + R_2 < I(U_1; Y_1 | T, X_1) + I(U_2; Y_2 | T, X_1) - I(U_1; U_2 | T, X_1), \\ R_0 + R_1 + R_2 < I(T, U_1; Y_1 | X_1) + I(U_2; Y_2 | T, X_1) - I(U_1; U_2 | T, X_1), \\ R_0 + R_1 + R_2 < I(U_1; Y_1 | T, X_1) + I(T, X_1, U_2; Y_2) - I(U_1; U_2 | T, X_1), \\ 2R_0 + R_1 + R_2 < I(T, U_1; Y_1 | X_1) + I(T, X_1, U_2; Y_2) - I(U_1; U_2 | T, X_1) \end{array} \right\}. \quad (4)$$

*Proof.* The proof of the achievability of $\mathcal{R}'_1$ is provided in Appendix A. The region $\mathcal{R}_1$ is obtained from $\mathcal{R}'_1$ by applying *Fourier-Motzkin elimination* (see, e.g., [17]) to eliminate $R'_1$ and $R'_2$ from the bounds in (3). □

The following lemma states a property [18] that we use to enlarge $\mathcal{R}_1$ in Theorem 2.

**Lemma 1.** *If the rate triple $(R_0, R_1, R_2)$ is achievable for the partially cooperative RBC, then the rate triple $(R_0 - \Delta_1 - \Delta_2, R_1 + \Delta_1, R_2 + \Delta_2)$ is also achievable, where $\Delta_1 \geq 0$, $\Delta_2 \geq 0$ and $\Delta_1 + \Delta_2 \leq R_0$.*

**Theorem 2.** *The following is an achievable region for the partially cooperative RBC:*

$$\mathcal{R}_2 = \bigcup_{p(x_1,t,u_1,u_2,x)p(y_1,y_2|x_1,x)} \left\{ \begin{array}{l} (R_0, R_1, R_2): \\ R_0 \geq 0, R_1 \geq 0, R_2 \geq 0, \\ R_0 + R_1 < I(T, U_1; Y_1 | X_1), \\ R_0 + R_2 < I(T, X_1, U_2; Y_2), \\ R_0 + R_1 + R_2 < I(T, U_1; Y_1 | X_1) + I(U_2; Y_2 | T, X_1) - I(U_1; U_2 | T, X_1), \\ R_0 + R_1 + R_2 < I(U_1; Y_1 | T, X_1) + I(T, X_1, U_2; Y_2) - I(U_1; U_2 | T, X_1), \\ 2R_0 + R_1 + R_2 < I(T, U_1; Y_1 | X_1) + I(T, X_1, U_2; Y_2) - I(U_1; U_2 | T, X_1) \end{array} \right\}. \quad (5)$$

*Proof.* We define

$$R'_0 = R_0 - \Delta_1 - \Delta_2, \quad R'_1 = R_1 + \Delta_1, \quad R'_2 = R_2 + \Delta_2 \quad (6)$$

where $\Delta_1 \geq 0$, $\Delta_2 \geq 0$, $\Delta_1 + \Delta_2 \leq R_0$. The region $\mathcal{R}_2$ is obtained by inserting (6) into the bounds (4) and applying *Fourier-Motzkin elimination* to eliminate $\Delta_1$ and $\Delta_2$ from those bounds. □

Note that for $\mathcal{R}_1$ and $\mathcal{R}'_1$ in Theorem 1, the relay decodes and forwards only the common message $W_0$, i.e., the relay does not help forward $W_2$. However, in Lemma 1 we move part of rate $R_0$ to $R_1$ or/and $R_2$, so the relay now forwards part of $W_2$.

The region $\mathcal{R}_2$ includes the following region that will be useful when we consider the semideterministic partially cooperative RBC in Section 5.

**Theorem 3.** *The following is an achievable region for the partially cooperative RBC:*

$$\mathcal{R}_3 = \bigcup_{p(x_1,t,u_1,u_2,x)p(y_1,y_2|x_1,x)} \left\{ \begin{array}{l} (R_0, R_1, R_2): \\ R_0 \geq 0, R_1 \geq 0, R_2 \geq 0, \\ R_0 < \min\{I(T; Y_1 | X_1), I(T, X_1; Y_2)\}, \\ R_0 + R_1 < I(T, U_1; Y_1 | X_1), \\ R_0 + R_2 < I(T, U_2, X_1; Y_2), \\ R_0 + R_1 + R_2 < I(T, U_1; Y_1 | X_1) + I(U_2; Y_2 | T, X_1) - I(U_1; U_2 | T, X_1), \\ R_0 + R_1 + R_2 < I(U_1; Y_1 | T, X_1) + I(T, X_1, U_2; Y_2) - I(U_1; U_2 | T, X_1) \end{array} \right\}. \quad (7)$$

*Proof.* The achievability of $\mathcal{R}_3$ follows from the achievability of $\mathcal{R}_2$ by observing that any rate triple that satisfies the bounds in $\mathcal{R}_3$ also satisfies the bounds in $\mathcal{R}_2$. In particular, the addition of the first bound on $R_0$ in $\mathcal{R}_3$ (the first term in the "min") with the last bound on $R_0 + R_1 + R_2$ in $\mathcal{R}_3$ implies the last bound in $\mathcal{R}_2$. An alternative proof of the achievability of $\mathcal{R}_3$ is given in Appendix C, where $\mathcal{R}_3$ is derived from the region $\mathcal{R}'_1$. This alternative proof exploits the geometric structure of $\mathcal{R}_3$. □

**Remark 1.** *The region $\mathcal{R}_3$ reduces to Marton's region for the broadcast channel (see [7, Theorem 2] and [19, P. 391, Prob. 10(c)]). This can be seen by setting $X_1 \equiv \phi$ in $\mathcal{R}_3$, i.e., disable the relay function for destination 1.*

**Remark 2.** *The region $\mathcal{R}_3$ reduces to the achievable rate of a partial decode-and-forward scheme for the relay channel given in [9]. This can be seen by setting $R_0 = 0$ and $R_1 = 0$, and choosing $U_1 = T$ and $U_2 = X$ in (7). The resulting bound is*

$$R_2 < \max \min \left\{ I(X_1, X; Y_2), I(T; Y_1|X_1) + I(X; Y_2|T, X_1) \right\} \tag{8}$$

*where the maximum is taken over the joint distributions $p(t, x_1, x) p(y_1, y_2 | x_1, x)$.*

We next provide an outer bound on the capacity region.

**Theorem 4.** *The following is an outer bound on the capacity region of the partially cooperative RBC:*

$$\bar{\mathcal{R}} = \bigcup_{p(t,u,x_1,x,y_1,y_2)} \left\{ \begin{array}{l} (R_0, R_1, R_2): \\ R_0 \geq 0, R_1 \geq 0, R_2 \geq 0, \\ R_0 \leq \min\left\{ I(T; Y_1|X_1), I(T, X_1; Y_2) \right\}, \\ R_0 + R_1 \leq I(X; Y_1|X_1), \\ R_0 + R_2 \leq \min\left\{ I(T, U, X_1; Y_2), I(X, X_1; Y_2) \right\}, \\ R_0 + R_1 + R_2 \leq I(T; Y_1|X_1) + I(X; Y_1|T, U, X_1) + I(U; Y_2|T, X_1), \\ R_0 + R_1 + R_2 \leq I(X; Y_1|T, U, X_1) + I(T, U, X_1; Y_2), \\ R_0 + R_1 + R_2 \leq I(X; Y_1, Y_2|X_1) \end{array} \right\} \tag{9}$$

*where the joint distribution $p(t, u, x_1, x, y_1, y_2)$ satisfies the Markov chain condition:*

$$(T, U) \to (X_1, X, Y_1) \to Y_2. \tag{10}$$

*Proof.* See Appendix B. □

**Remark 3.** *The region $\bar{\mathcal{R}}$ is convex. In fact, the random variable $Q$ defined at the end of Appendix B can be viewed as a time-sharing random variable.*

**Remark 4.** *The joint distribution $p(t, u, x_1, x, y_1, y_2)$ of the random variables in Theorem 4 does not necessarily satisfy the Markov chain condition:*

$$(T, U) \to (X_1, X) \to (Y_1, Y_2). \tag{11}$$

**Remark 5.** *The outer bound $\bar{\mathcal{R}}$ is at least as good as cut-set bounds, and sometimes tighter.*

## 3.1 New Bounds for Broadcast Channels

The capacity region of the discrete memoryless broadcast channel is still unknown. Inner and outer bounds have been obtained in, e.g., [14, 7, 18], and e.g., [7, 8], respectively. We provide new inner and outer bounds derived from Theorems 2 and 4, respectively.

**Theorem 5.** *The following is an inner bound on the capacity region of the broadcast channel:*

$$\mathcal{R}_{2,BC} = \bigcup_{p(t,u_1,u_2,x)p(y_1,y_2|x)} \left\{ \begin{array}{l} (R_0, R_1, R_2) : R_0 \geq 0, R_1 \geq 0, R_2 \geq 0, \\ R_0 + R_1 < I(T, U_1; Y_1), \\ R_0 + R_2 < I(T, U_2; Y_2), \\ R_0 + R_1 + R_2 < I(T, U_1; Y_1) + I(U_2; Y_2|T) - I(U_1; U_2|T), \\ R_0 + R_1 + R_2 < I(U_1; Y_1|T) + I(T, U_2; Y_2) - I(U_1; U_2|T), \\ 2R_0 + R_1 + R_2 < I(T, U_1; Y_1) + I(T, U_2; Y_2) - I(U_1; U_2|T) \end{array} \right\}. \qquad (12)$$

*Proof.* The proof follows by setting $X_1 = \phi$ in Theorem 2. □

Among the inner bounds that have been obtained for the broadcast channel so far, Marton's region given in [7, Theorem 2] and [19, P. 391, Prob. 10(c)] is the largest, and is sometimes conjectured to be the capacity region. Note that Marton's region can be derived from $\mathcal{R}_3$ in Theorem 3 (see Remark 1). Comparing Theorem 5 with Marton's region in [19, P. 391, Prob. 10(c)], we can make the following formal remark.

**Remark 6.** *The inner bound $\mathcal{R}_{2,BC}$ includes Marton's region. We do not know if $\mathcal{R}_{2,BC}$ is strictly larger, however.*

The following outer bound $\bar{\mathcal{R}}_{BC}$ for the broadcast channel can be specialized from the outer bound $\bar{\mathcal{R}}$ in Theorem 4.

**Theorem 6.** *The following is an outer bound on the capacity region of the broadcast channel:*

$$\bar{\mathcal{R}}_{BC} = \bigcup_{p(t,u,v,x)p(y_1,y_2|x)} \left\{ \begin{array}{l} (R_0, R_1, R_2) : R_0 \geq 0, R_1 \geq 0, R_2 \geq 0, \\ R_0 \leq \min \{I(T; Y_1), I(T; Y_2)\}, \\ R_0 + R_1 \leq I(X; Y_1), \\ R_0 + R_2 \leq I(T, U; Y_2), \\ R_0 + R_1 + R_2 \leq I(T; Y_1) + I(X; Y_1|T, U) + I(U; Y_2|T), \\ R_0 + R_1 + R_2 \leq I(T, U; Y_2) + I(X; Y_1|T, U), \\ R_0 + R_1 \leq I(T, V; Y_1), \\ R_0 + R_2 \leq I(X; Y_2), \\ R_0 + R_1 + R_2 \leq I(T, V; Y_1) + I(X; Y_2|T, V), \\ R_0 + R_1 + R_2 \leq I(T; Y_2) + I(V; Y_1|T) + I(X; Y_2|T, V) \end{array} \right\}. \qquad (13)$$

*Proof.* The first five nontrivial bounds can be obtained from Theorem 4 by setting $X_1 \equiv \phi$. The last four bounds can be obtained by switching the roles of destinations 1 and 2. Note that the joint distribution $p(t, u, v, x, y_1, y_2)$ satisfies the Markov chain condition:

$$(T, U, V) \to X \to (Y_1, Y_2). \tag{14}$$

One can check this condition by referring to the proof of Theorem 4 given in Appendix B. $\square$

**Remark 7.** *The outer bound $\bar{\mathcal{R}}_{BC}$ reduces to the outer bound in [7, Theorem 5] for the broadcast channel with only private message sets. It also reduces to the outer bound in [8] for the semideterministic broadcast channel with both a common message and two private messages. Note that our proof is simpler than the proof given in [8] that was based on a recursive approach.*

**Remark 8.** *The outer bound $\bar{\mathcal{R}}_{BC}$ gives the capacity region for the more capable broadcast channel studied in [20].*

## 4 Partially Cooperative RBCs with Degraded Message Sets

In this section, we consider the *partially cooperative RBC with degraded message sets* where $R_2 = 0$ (see Fig. 3). We derive the following inner and outer bounds on the capacity region.

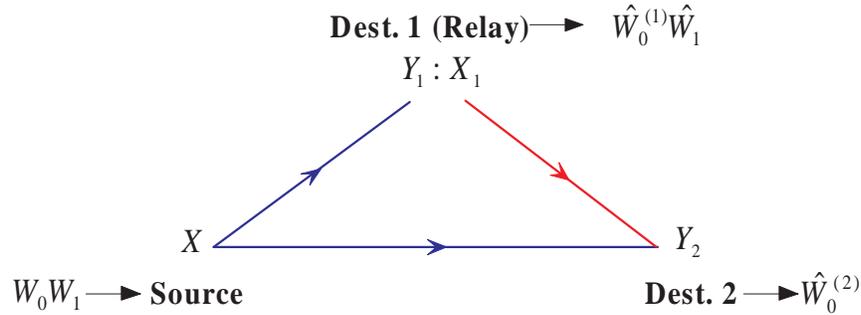

Figure 3: Partially cooperative RBC with degraded message sets

**Theorem 7.** *The capacity region $\mathcal{C}_{dm}$ of the partially cooperative RBC with degraded message sets satisfies*

$$\begin{aligned}
\mathcal{C}_{dm} &\supseteq \bigcup_{p(t,x_1,x,y_1,y_2)=p(t,x_1,x)p(y_1,y_2|x,x_1)} \mathcal{R}\Big(p(t, x_1, x, y_1, y_2)\Big) \\
\mathcal{C}_{dm} &\subseteq \bigcup_{p(t,x_1,x,y_1,y_2)=p(t,x_1,x,y_1)p(y_2|y_1,x,x_1)} \mathcal{R}\Big(p(t, x_1, x, y_1, y_2)\Big)
\end{aligned} \tag{15}$$

*where*

$$\mathcal{R}\Big(p(t, x_1, x, y_1, y_2)\Big) = \left\{ \begin{array}{l} (R_0, R_1): \\ R_0 \geq 0, R_1 \geq 0, \\ R_0 < I(T, X_1; Y_2), \\ R_0 + R_1 < I(X; Y_1|X_1), \\ R_0 + R_1 < I(X; Y_1|T, X_1) + I(T, X_1; Y_2) \end{array} \right\}. \quad (16)$$

*Proof.* The inner bound follows from Theorem 2 by setting $R_2 = 0$ and choosing $U_1 = X$ and $U_2 = T$. To derive the outer bound, we use Theorem 4. The first and second bounds in (16) follow from the respective first and second bounds in (9). The third bound in (16) follows by setting $R_2 = 0$ in Theorem 4 and writing $(T, U)$ as $T$ for the fifth bound in (9). □

**Remark 9.** *The inner and outer bounds given in Theorem 7 have the same form except that the joint distribution $p(t, x_1, x, y_1, y_2)$ satisfies different Markov chains.*

## 5 Semideterministic Partially Cooperative RBCs

In this section, we specialize our theory to semideterministic partially cooperative RBCs (see Fig. 4).

**Definition 1.** *A partially cooperative RBC is* semideterministic *if the transition probability distribution $p(y_1|x, x_1)$ takes on the values 0 or 1 only. The channel is* deterministic *if $p(y_1, y_2|x, x_1)$ takes on the values 0 or 1 only.*

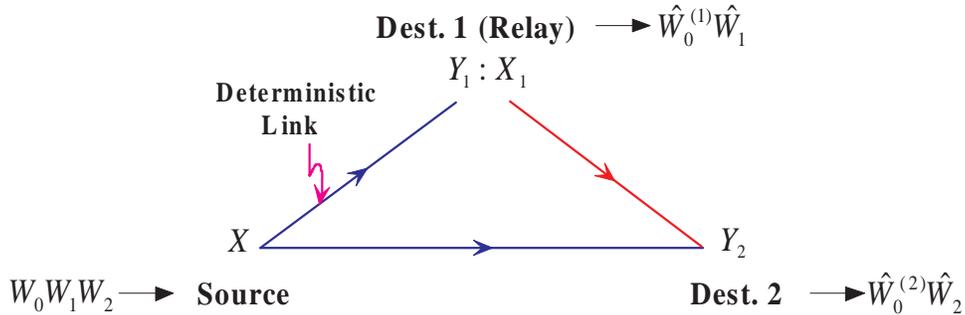

Figure 4: Semideterministic partially cooperative RBC

**Theorem 8.** *The capacity region of the semideterministic partially cooperative RBC is*

$$\mathcal{C}_{sd} = \bigcup_{p(t,u,x_1,x)p(y_1,y_2|x,x_1)} \left\{ \begin{array}{l} (R_0, R_1, R_2) : \\ R_0 \leq \min\left\{I(T;Y_1|X_1), I(T,X_1;Y_2)\right\}, \\ R_0 + R_1 \leq H(Y_1|X_1), \\ R_0 + R_2 \leq I(T,U,X_1;Y_2), \\ R_0 + R_1 + R_2 \leq I(T;Y_1|X_1) + H(Y_1|T,U,X_1) + I(U;Y_2|T,X_1) \\ R_0 + R_1 + R_2 \leq H(Y_1|T,U,X_1) + I(T,X_1,U;Y_2) \end{array} \right\}. \quad (17)$$

*Proof.* Achievability follows from Theorem 3 by setting $U_1 = Y_1$, redefining $U_2 = U$, and using $H(Y_1|X_1, X) = 0$. The converse follows from Theorem 4 by using $H(Y_1|X_1, X) = 0$. We are left to show that the joint distribution $p(t, u, x_1, x, y_1, y_2)$ for the outer bound satisfies the Markov chain condition:

$$(T, U) \to (X_1, X) \to (Y_1, Y_2). \quad (18)$$

This condition follows from the proof of Theorem 4 given in Appendix B. □

**Remark 10.** *Theorem 8 reduces to the capacity region of the semideterministic broadcast channel given in [8] by setting $X_1 = \phi$.*

**Remark 11.** *Theorem 8 reduces to the capacity of the semideterministic relay channel given in [9] by setting $R_0 = 0$, $R_1 = 0$, $T = Y_1$, and $U = X$.*

It is now easy to derive the following capacity region from Theorem 8.

**Corollary 1.** *The capacity region of the deterministic partially cooperative RBC is*

$$\mathcal{C}_d = \bigcup_{p(t,x_1,x)p(y_1,y_2|x,x_1)} \left\{ \begin{array}{l} (R_0, R_1, R_2) : \\ R_0 \geq 0, R_1 \geq 0, R_2 \geq 0, \\ R_0 \leq \min\left\{I(T;Y_1|X_1), I(T,X_1;Y_2)\right\}, \\ R_0 + R_1 \leq H(Y_1|X_1), \\ R_0 + R_2 \leq H(Y_2), \\ R_0 + R_1 + R_2 \leq I(T;Y_1|X_1) + H(Y_1,Y_2|T,X_1) \\ R_0 + R_1 + R_2 \leq I(T,X_1;Y_2) + H(Y_1,Y_2|T,X_1) \end{array} \right\}. \quad (19)$$

*Proof.* Achievability follows by setting $U = Y_2$ in Theorem 8. To prove the converse, consider Theorem 4 and note that the first two bounds in (19) follow from the first two bounds in (9) by using $H(Y_1|X_1, X) = 0$. The third bound in (19) follows from the third bound in (9) by using $H(Y_2|X_1, X) = 0$. We next use the fourth bound in (9) and obtain the fourth bound in (19):

$$\begin{aligned}
R_0 + R_1 + R_2 &\leq I(T;Y_1|X_1) + I(X;Y_1|T,U,X_1) + I(U;Y_2|T,X_1) \\
&\leq I(T;Y_1|X_1) + I(X;Y_1,Y_2|T,U,X_1) + I(U;Y_1,Y_2|T,X_1) \\
&= I(T;Y_1|X_1) + I(X,U;Y_1,Y_2|T,X_1) \\
&= I(T;Y_1|X_1) + H(Y_1,Y_2|T,X_1)
\end{aligned} \quad (20)$$

where the last step follows by using $H(Y_1, Y_2|X_1, X) = 0$.

Finally, we use the fifth bound in (9) and obtain the fifth bound in (19):

$$\begin{aligned}
R_0 + R_1 + R_2 &\leq I(X;Y_1|T,U,X_1) + I(T,U,X_1;Y_2) \\
&\leq I(X;Y_1|T,U,X_1) + I(U;Y_2|T,X_1) + I(T,X_1;Y_2) \\
&\leq I(X;Y_1,Y_2|T,U,X_1) + I(U;Y_1,Y_2|T,X_1) + I(T,X_1;Y_2) \\
&\leq I(X,U;Y_1,Y_2|T,X_1) + I(T,X_1;Y_2) \\
&\leq H(Y_1,Y_2|T,X_1) + I(T,X_1;Y_2).
\end{aligned} \quad (21)$$

□

Corollary 1 has an especially simple form when $R_0 = 0$.

**Corollary 2.** *The capacity region of the deterministic partially cooperative RBC with $R_0 = 0$ is*

$$\mathcal{C}_d = \bigcup_{\substack{p(x_1, x) \\ p(y_1, y_2|x, x_1)}} \left\{ \begin{array}{l} (R_1, R_2): \\ R_1 \geq 0, R_2 \geq 0, \\ R_1 \leq H(Y_1|X_1), \\ R_2 \leq H(Y_2), \\ R_1 + R_2 \leq H(Y_1, Y_2|X_1) \end{array} \right\}. \quad (22)$$

*Proof.* Achievability follows by setting $T = \phi$ and $R_0 = 0$ in Corollary 1. The converse follows from cut-set bounds. □

We now consider an example channel that we call the Blackwell partially cooperative RBC.

**Example 1.** *Suppose the channel from the source to destinations 1 and 2 is the Blackwell broadcast channel [21, 22] with the channel depicted in Fig. 5. Suppose further that the channel from destination 1 to destination 2 is orthogonal to the channel from the source to destinations 1 and 2, and is deterministic with capacity $r$ (see Fig. 5). We refer to this channel as the Blackwell partially cooperative RBC.*

Suppose that $R_0 = 0$. From Corollary 2, we obtain the following bounds:

$$\begin{aligned}
R_1 &\leq H(Y_1) \\
R_2 &\leq H(Y_2) \overset{(a)}{\leq} H(Y_{21}) + H(Y_{22}) \\
R_1 + R_2 &\leq H(Y_1, Y_2|X_1) = H(Y_1, Y_{21}, Y_{22}|X_1) \\
&= H(Y_1, Y_{21}|X_1) \overset{(b)}{\leq} H(Y_1, Y_{21})
\end{aligned} \quad (23)$$

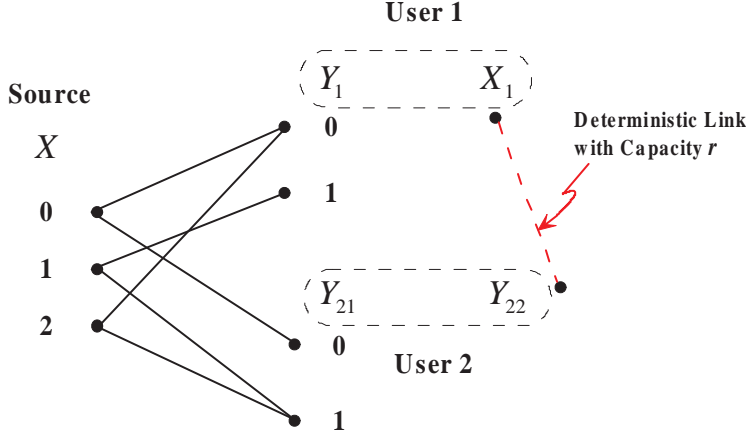

Figure 5: Blackwell partially cooperative RBC

where $(a)$ and $(b)$ are equalities if and only if $X$ is independent of $X_1$. It is clear that the input distributions $p(x, x_1)$ that achieve capacity boundary points have $X$ independent of $X_1$. We parameterize the distribution of $X$ as follows:

$$Pr\{X = 0\} = \alpha, \qquad Pr\{X = 1\} = \beta, \qquad Pr\{X = 2\} = 1 - \alpha - \beta \qquad (24)$$

where $\alpha \geq 0$, $\beta \geq 0$ and $\alpha + \beta \leq 1$. Inserting (24) into (23), we obtain the capacity region for the Blackwell partially cooperative RBC:

$$\mathcal{C}_{Bw,RBC} = \bigcup_{\substack{\alpha \geq 0, \beta \geq 0, \\ \alpha + \beta \leq 1}} \left\{ \begin{array}{l} (R_1, R_2): \\ R_1 \geq 0, R_2 \geq 0, \\ R_1 \leq h(\beta), \\ R_2 \leq r + h(\alpha), \\ R_1 + R_2 \leq -\alpha \log \alpha - \beta \log \beta \\ \qquad\qquad -(1 - \alpha - \beta) \log(1 - \alpha - \beta) \end{array} \right\} \qquad (25)$$

where $h(x) := -x \log x - (1-x) \log(1-x)$.

If destination 1 does not relay information for destination 2, the RBC reduces to the Blackwell broadcast channel. In this case, the region (25) reduces to (see [22])

$$\mathcal{C}_{Bw,BC} = \bigcup_{\substack{\alpha \geq 0, \beta \geq 0, \\ \alpha + \beta \leq 1}} \left\{ \begin{array}{l} (R_1, R_2): \\ R_1 \geq 0, R_2 \geq 0, \\ R_1 \leq h(\beta), \\ R_2 \leq h(\alpha), \\ R_1 + R_2 \leq -\alpha \log \alpha - \beta \log \beta \\ \qquad\qquad -(1 - \alpha - \beta) \log(1 - \alpha - \beta) \end{array} \right\}. \qquad (26)$$

We compare the capacity region (25) for different $r$ in Fig. 6 (the capacity region of the Blackwell broadcast channel (26) is given by the dashed curve labelled $r = 0$). From this figure, it is clear that relaying improves rates by enlarging the bound on $R_2$ in (25). The capacity region grows as $r$ increases because the relay is able to transmit more information to destination 2.

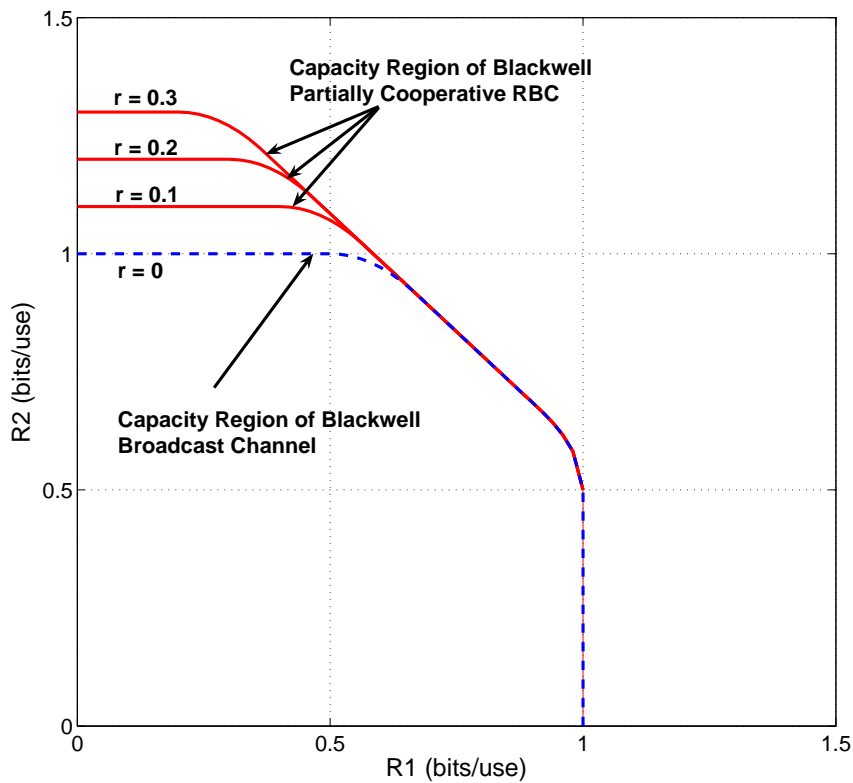

Figure 6: Capacity region of the Blackwell partially cooperative RBC

## 6 Orthogonal Partially Cooperative RBCs

In this section, we study the *orthogonal partially cooperative RBC* where the relay (destination 1) has the practical constraint that it must transmit and receive in two orthogonal channels. The channel model is illustrated in Fig. 7 and is defined as follows.

**Definition 2.** *An orthogonal partially cooperative RBC consists of two source input alphabets $\mathcal{X}_R$ and $\mathcal{X}_D$, a relay input alphabet $\mathcal{X}_1$, two channel output alphabets $\mathcal{Y}_1$ and $\mathcal{Y}_2$, and a transition probability distribution*

$$p(y_1, y_2 | x_R, x_D, x_1) = p(y_1 | x_R, x_1) p(y_2 | x_D, x_1). \tag{27}$$

Note that this model reduces to the relay channel with orthogonal components studied in [10] if the source has only a private message $W_2$ for destination 2. We have the following capacity theorem.

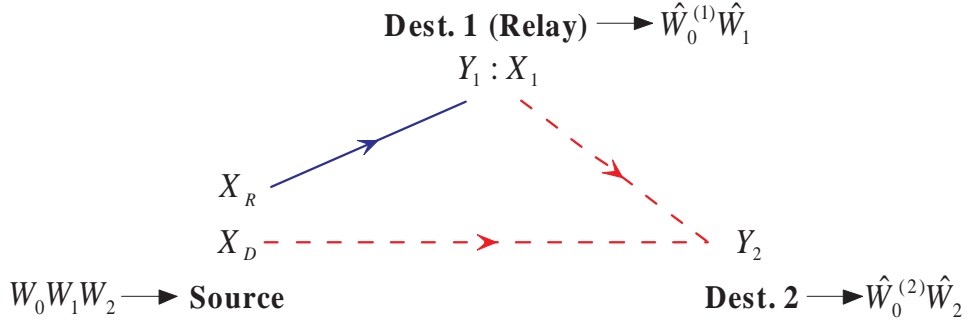

Figure 7: Orthogonal partially cooperative RBC

**Theorem 9.** *The capacity region of the orthogonal partially cooperative RBC is*

$$\mathcal{C}_{or} = \bigcup_{\substack{p(x_1)p(x_R|x_1)p(x_D|x_1) \\ p(y_1|x_R,x_1)p(y_2|x_D,x_1)}} \left\{ \begin{array}{l} (R_0, R_1, R_2): \\ R_0 \geq 0, R_1 \geq 0, R_2 \geq 0, \\ R_0 + R_1 \leq I(X_R; Y_1|X_1), \\ R_0 + R_2 \leq I(X_D, X_1; Y_2), \\ R_0 + R_1 + R_2 \leq I(X_R; Y_1|X_1) + I(X_D; Y_2|X_1) \end{array} \right\}. \quad (28)$$

*Proof.* To prove achievability, consider Theorem 2 and set $T = \phi$, $U_1 = X_R$, and $U_2 = X_D$. We further choose $p(x_1, x_R, x_D) = p(x_1)p(x_R|x_1)p(x_D|x_1)$. The first, second and third bounds in (5) provide the achievable region, and the other two bounds in (5) are implied by these three bounds. The converse follows from cut-set bounds. □

An alternative proof of achievability is provided in Appendix D. From this proof, we see that the source does not apply superposition encoding or binning, and that block Markov encoding is not necessary. Hence, the code construction and encoding is much simpler for the orthogonal partially cooperative RBC than for a general partially cooperative RBC.

**Remark 12.** *Theorem 9 reduces to the capacity of the relay channel with orthogonal components given in [10]. The result is*

$$R_2 \leq \min\{I(X_D, X_1; Y_2), I(X_R; Y_1|X_1) + I(X_D; Y_2|X_1)\}. \quad (29)$$

We next consider the Gaussian orthogonal partially cooperative RBC with channel outputs

$$\begin{aligned} Y_1 &= X_R + Z_1 \\ Y_2 &= X_D + X_1 + Z_2 \end{aligned} \quad (30)$$

where $Z_1$ and $Z_2$ are independent, zero mean, Gaussian random variables with variances $N_1$ and $N_2$, respectively. The channel input sequences $\{(x_{R,i}, x_{D,i})\}$ and $\{x_{1,i}\}$ are subject to the following average power constraints:

$$\frac{1}{n}\sum_{i=1}^{n} \mathsf{E}\left[X_{R,i}^2 + X_{D,i}^2\right] \leq P, \quad \text{and} \quad \frac{1}{n}\sum_{i=1}^{n} \mathsf{E}\left[X_{1,i}^2\right] \leq P_1. \quad (31)$$

The capacity region for this channel is as follows.

**Corollary 3.** *The capacity region for the Gaussian orthogonal partially cooperative RBC is*

$$\mathcal{C}_{G,or} = \bigcup_{\alpha,\beta \in [0,1]} \left\{ \begin{array}{l} (R_0, R_1, R_2): \\ R_0 \geq 0, R_1 \geq 0, R_2 \geq 0, \\ R_0 + R_1 \leq \mathcal{C}\left(\dfrac{\alpha P}{N_1}\right), \\ R_0 + R_2 \leq \mathcal{C}\left(\dfrac{P_1 + \bar{\alpha} P + 2\sqrt{\beta \bar{\alpha} P P_1}}{N_2}\right), \\ R_0 + R_1 + R_2 \leq \mathcal{C}\left(\dfrac{\alpha P}{N_1}\right) + \mathcal{C}\left(\dfrac{\beta \bar{\alpha} P}{N_2}\right) \end{array} \right\} \quad (32)$$

where $\mathcal{C}(x) = \frac{1}{2}\log(1+x)$.

*Proof.* The proof is a simple extension of Theorem 9 and is omitted. □

# 7 Parallel Partially Cooperative RBCs with Unmatched Degraded Subchannels

In this section, we consider the parallel partially cooperative RBC with unmatched degraded subchannels. This channel includes two partially cooperative RBCs (subchannels I and II) with communication links in two channels orthogonal to each other. Moreover, the output of destination 2 is a degraded version of the output of destination 1 in subchannel I, and the output of destination 1 is a degraded version of the output of destination 2 in subchannel II. The channel model is illustrated in Fig. 8 and is defined as follows.

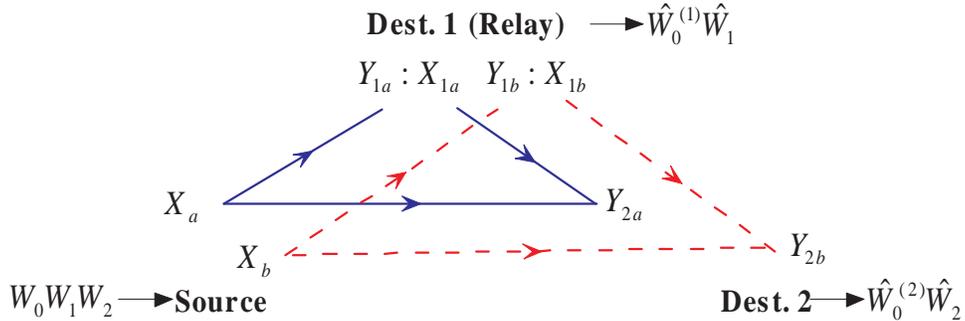

Figure 8: Parallel partially cooperative RBC with unmatched degraded subchannels

**Definition 3.** *A parallel partially cooperative RBC with unmatched degraded subchannels has a vector source input alphabet* $(\mathcal{X}_a, \mathcal{X}_b)$, *a vector relay input alphabet* $(\mathcal{X}_{1a}, \mathcal{X}_{1b})$, *two* *vector output alphabets* $(\mathcal{Y}_{1a}, \mathcal{Y}_{1b})$ *and* $(\mathcal{Y}_{2a}, \mathcal{Y}_{2b})$, *and a transition probability distribution:*

$$p(y_{1a}, y_{2a}|x_a, x_{1a})p(y_{1b}, y_{2b}|x_b, x_{1b})$$

*that satisfies the following two degradedness conditions:*

$$\begin{aligned} p(y_{1a}, y_{2a}|x_a, x_{1a}) &= p(y_{1a}|x_a, x_{1a})p(y_{2a}|y_{1a}, x_{1a}) \\ p(y_{1b}, y_{2b}|x_b, x_{1b}) &= p(y_{2b}|x_b, x_{1b})p(y_{1b}|y_{2b}, x_{1b}). \end{aligned} \quad (33)$$

Note that the parallel partially cooperative RBC with unmatched degraded subchannels is not a degraded partially cooperative RBC, although both subchannels are degraded. We obtain the following bounds on the capacity region.

**Theorem 10.** *An achievable region for the parallel partially cooperative RBC with unmatched degraded subchannels is*

$$\mathcal{R}_{para} = \bigcup_{p(t_a, x_a, x_{1a})p(t_b, x_b, x_{1b})p(y_{1a}, y_{2a}|x_a, x_{1a})p(y_{1b}, y_{2b}|x_b, x_{1b})} \left\{ \begin{array}{l} (R_0, R_1, R_2): \\ R_0 \geq 0, R_1 \geq 0, R_2 \geq 0, \\ R_0 < I(T_a; Y_{1a}|X_{1a}) + I(T_b; Y_{1b}|X_{1b}), \\ R_0 < I(T_a, X_{1a}; Y_{2a}) + I(T_b, X_{1b}; Y_{2b}), \\ R_0 + R_1 < I(X_a; Y_{1a}|X_{1a}) + I(T_b; Y_{1b}|X_{1b}), \\ R_0 + R_2 < I(T_a, X_{1a}; Y_{2a}) + I(X_b, X_{1b}; Y_{2b}), \\ R_0 + R_1 + R_2 < I(X_a; Y_{1a}|X_{1a}) + I(T_b; Y_{1b}|X_{1b}) + I(X_b; Y_{2b}|T_b, X_{1b}), \\ R_0 + R_1 + R_2 < I(X_a; Y_{1a}|T_a, X_{1a}) + I(T_a, X_{1a}; Y_{2a}) + I(X_b, X_{1b}; Y_{2b}) \end{array} \right\}. \quad (34)$$

*An outer bound on the capacity region is*

$$\bar{\mathcal{R}}_{para} = \bigcup_{p(x_a, x_{1a})p(x_b, x_{1b})p(y_{1a}, y_{2a}|x_a, x_{1a})p(y_{1b}, y_{2b}|x_b, x_{1b})} \left\{ \begin{array}{l} (R_0, R_1, R_2): \\ R_0 \geq 0, R_1 \geq 0, R_2 \geq 0, \\ R_0 + R_1 \leq I(X_a; Y_{1a}|X_{1a}) + I(X_b; Y_{1b}|X_{1b}), \\ R_0 + R_2 \leq I(X_a, X_{1a}; Y_{2a}) + I(X_b, X_{1b}; Y_{2b}), \\ R_0 + R_1 + R_2 \leq I(X_a; Y_{1a}|X_{1a}) + I(X_b; Y_{2b}|X_{1b}) \end{array} \right\}. \quad (35)$$

*Proof.* The inner bound follows by choosing $Y_1 = (Y_{1a}, Y_{1b})$, $Y_2 = (Y_{2a}, Y_{2b})$, $T = (T_a, T_b)$, $U_1 = X_a$ and $U_2 = X_b$ in Theorem 3, and by assuming that $(T_a, X_a, X_{1a})$ is independent of $(T_b, X_b, X_{1b})$. The outer bound follows from cut-set bounds and the degradedness conditions (33). □

**Remark 13.** *The region $\mathcal{R}_{para}$ in Theorem 10 reduces to the capacity region of the product of two unmatched degraded broadcast channels given in [12].*

In general, the inner and outer bounds given in Theorem 10 do not match. We next consider three cases of this channel where the source has different message sets for the two destinations. Case 1 has $R_2 = 0$, i.e., the source has a common message $W_0$ for both destinations and a private message $W_1$ for destination 1. We obtain the following capacity region for this case.

**Theorem 11.** *For the parallel partially cooperative RBC with unmatched degraded subchannels where $R_2 = 0$, the capacity region is*

$$\mathcal{C}_{para}^{I} = \bigcup_{p(t,x_a,x_{1a})p(x_b,x_{1b})p(y_{1a},y_{2a}|x_a,x_{1a})p(y_{1b},y_{2b}|x_b,x_{1b})} \left\{ \begin{array}{l} (R_0, R_1): \\ R_0 \geq 0, R_1 \geq 0, \\ R_0 \leq I(T, X_{1a}; Y_{2a}) + I(X_b, X_{1b}; Y_{2b}), \\ R_0 + R_1 \leq I(X_a; Y_{1a}|X_{1a}) + I(X_b; Y_{1b}|X_{1b}), \\ R_0 + R_1 + R_2 \leq I(X_a; Y_{1a}|T, X_{1a}) + I(T, X_{1a}; Y_{2a}) + I(X_b, X_{1b}; Y_{2b}) \end{array} \right\}. \quad (36)$$

*Proof.* To prove achievability, we use (16) in Theorem 7 with $T = (T_a, X_b)$, $X = (X_a, X_b)$, $Y_1 = (Y_{1a}, Y_{1b})$ and $Y_2 = (Y_{2a}, Y_{2b})$, and assume $(T_a, X_a, X_{1a})$ is independent of $(X_b, X_{1b})$. For notational convenience, we finally replace $T_a$ with $T$. The converse is given in Appendix E. □

We now consider case 2, where $R_0 = 0$ and $R_1 = 0$, i.e., the source has only a private message $W_2$ for destination 2. The channel in this case reduces to the parallel relay channel with unmatched degraded subchannels. Note that this relay channel is not degraded as defined in [3]. We obtain the following capacity result.

**Theorem 12.** *The capacity of the parallel relay channel with unmatched degraded subchannels is*

$$C_{para}^{II} = \max \min\{I(X_a, X_{1a}; Y_{2a}) + I(X_b, X_{1b}; Y_{2b}), \\ I(X_a; Y_{1a}|X_{1a}) + I(X_b; Y_{2b}|X_{1b})\}. \quad (37)$$

*where the maximum is taken over all joint distributions*

$$p(x_a, x_{1a})p(x_b, x_{1b})p(y_{1a}, y_{2a}|x_a, x_{1a})p(y_{1b}, y_{2b}|x_b, x_{1b}).$$

*Proof.* Achievability follows by setting $R_0 = 0$ and $R_1 = 0$ and choosing $T_a = X_a$, $T_b = \phi$ in (34) in Theorem 10. The converse follows by setting $R_0 = 0$ and $R_1 = 0$ in the last two bounds in (35). □

**Remark 14.** *Theorem 12 establishes the capacity for a new class of relay channels.*

Note that the capacity $C$ of the point-to-point parallel channel with two subchannels is $C = C_a + C_b$ if the capacities of the two subchannels are $C_a$ and $C_b$, respectively. However, this is not true for the parallel relay channel. To see this, observe that the capacities of the two degraded relay subchannels are (see [3]):

$$C_a = \max_{p(x_a, x_{1a})p(y_{1a}, y_{2a}|x_a, x_{1a})} \min\{I(X_a, X_{1a}; Y_{2a}), I(X_a; Y_{1a}|X_{1a})\}$$
$$C_b = \max_{p(x_b, x_{1b})p(y_{1b}, y_{2b}|x_b, x_{1b})} I(X_b; Y_{2b}|X_{1b}). \tag{38}$$

But the sum of $C_a$ and $C_b$ in (38) is

$$\begin{aligned} C_a + C_b &= \max \min\{I(X_a, X_{1a}; Y_{2a}) + I(X_b; Y_{2b}|X_{1b}), \\ &\qquad I(X_a; Y_{1a}|X_{1a}) + I(X_b; Y_{2b}|X_{1b})\} \\ &\leq C_{para}^{II} \end{aligned} \tag{39}$$

where the maximum is taken over all joint distributions

$$p(x_a, x_{1a})p(x_b, x_{1b})p(y_{1a}, y_{2a}|x_a, x_{1a})p(y_{1b}, y_{2b}|x_b, x_{1b}).$$

**Remark 15.** *The capacity of the parallel relay channel with two subchannels can be larger than the sum of the capacities of the two subchannels.*

Intuitively, Remark 15 follows because information transmitted over the two subchannels may not be independent as in the point-to-point parallel channel with two subchannels. As we show in the following example, information forwarded from the source to the relay in one subchannel may be forwarded from the relay to the destination in the other subchannel.

**Example 2.** *Consider the two subchannels shown in Fig. 9 and suppose all channel input and output alphabets are $\{0, 1\}$.*

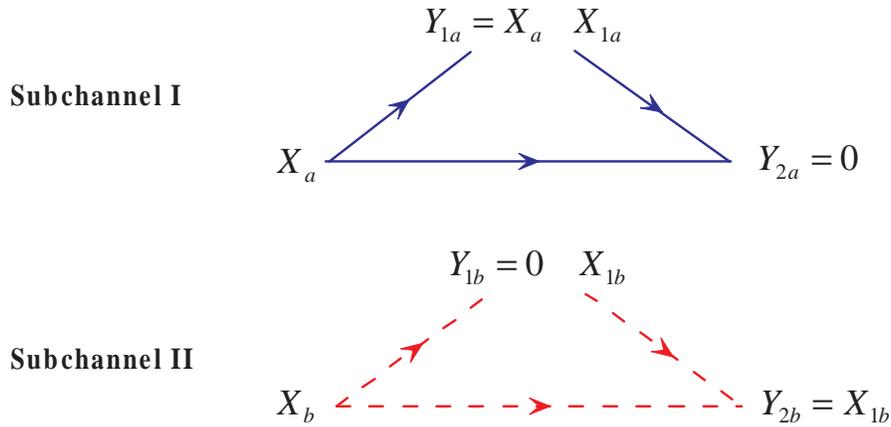

Figure 9: An example of parallel relay channel with unmatched degraded subchannels

One can check that subchannel I is degraded with capacity $C_a = 0$, and subchannel II is reversely degraded with capacity $C_b = 0$. However, by (37) the capacity of the parallel relay channel is $C_{para}^{II} = 1$ that is larger than $C_a + C_b$. This capacity is achieved by having the source node forward 1 bit per channel use to the relay over subchannel I, and the relay forward 1 bit per channel use to the destination node over subchannel II.

Finally, consider case 3 where $R_0 = 0$, i.e., the source has private messages $W_1$ and $W_2$ for destinations 1 and 2, respectively.

**Corollary 4.** *For the parallel partially cooperative RBC with unmatched degraded subchannels where $R_0 = 0$, an achievable region is*

$$\mathcal{R}_{para}^{III} = \bigcup_{p(t_a, x_a, x_{1,a}) p(t_b, x_b, x_{1,b}) p(y_{1,a}, y_{2,a}|x_a, x_{1,a}) p(y_{1,b}, y_{2,b}|x_b, x_{1,b})} \left\{ \begin{array}{l} (R_0, R_1, R_2): \\ R_0 \geq 0, R_1 \geq 0, R_2 \geq 0, \\ R_1 < I(X_a; Y_{1a}|X_{1a}) + I(T_b; Y_{1b}|X_{1b}), \\ R_2 < I(T_a, X_{1a}; Y_{2a}) + I(X_b, X_{1b}; Y_{2b}), \\ R_1 + R_2 < I(X_a; Y_{1a}|X_{1a}) + I(T_b; Y_{1b}|X_{1b}) + I(X_b; Y_{2b}|T_b, X_{1b}), \\ R_1 + R_2 < I(X_a; Y_{1a}|T_a, X_{1a}) + I(T_a, X_{1a}; Y_{2a}) + I(X_b, X_{1b}; Y_{2b}) \end{array} \right\}. \quad (40)$$

*Proof.* The result follows directly from Theorem 10. □

Note that the two subchannels in Corollary 4 are the degraded RBC with the capacity region given in [1] and the reversely degraded RBC with the capacity region given in [4]. The capacity regions of the two subchannels are as follows:

$$\mathcal{C}_a = \bigcup_{\substack{p(t_a, x_a, x_{1a}) \\ p(y_{1a}, y_{2a}|x_a, x_{1a})}} \left\{ \begin{array}{l} (R_1, R_2): \\ R_1 \geq 0, R_2 \geq 0, \\ R_1 \leq I(X_a; Y_{1a}|T_a, X_{1a}) \\ R_2 \leq \min\{I(T_a, X_{1a}; Y_{2a}), I(T_a; Y_{1a}|X_{1a})\} \end{array} \right\} \quad (41)$$

and

$$\mathcal{C}_b = \bigcup_{\substack{p(t_b, x_b, x_{1b}) \\ p(y_{1b}, y_{2b}|x_b, x_{1b})}} \left\{ \begin{array}{l} (R_1, R_2): \\ R_1 \geq 0, R_2 \geq 0, \\ R_1 \leq I(T_b; Y_{1b}|X_{1b}) \\ R_2 \leq I(X_b; Y_{2b}|T_b, X_{1b}) \end{array} \right\}. \quad (42)$$

The Minkowski sum of the preceding two capacity regions is given by

$$\mathcal{C}_a + \mathcal{C}_b := \left\{ \begin{array}{l} (R_1, R_2): R_1 = R_{1a} + R_{1b}, R_2 = R_{2a} + R_{2b}, \\ \text{where } (R_{1a}, R_{2a}) \in \mathcal{C}_a, (R_{1b}, R_{2b}) \in \mathcal{C}_b \end{array} \right\} \quad (43)$$

From the preceding relay channel example, it is clear that, in general, we have

$$\mathcal{C}_a + \mathcal{C}_b \subset \mathcal{R}_{para}^{III}. \quad (44)$$

We hence make the following remark.

**Remark 16.** *The achievable region of the parallel partially cooperative RBC with unmatched degraded subchannels where $R_0 = 0$ can be larger than the Minkowski sum of the capacity regions of the two subchannels. This is in contrast to the parallel broadcast channel with unmatched degraded subchannels and $R_0 = 0$, where the capacity region equals the Minkowski sum of the capacity regions of the two subchannels [11, 12].*

## 8 Conclusions

We derived new inner and outer bounds on the capacity region of the discrete memoryless partially cooperative RBC. The inner bounds are based on the source using superposition encoding and binning, and the relay using decode-and-forward for the common message and partial decode-and-forward for the private message for destination 2. Our outer bound on the capacity region uses a more general and simpler approach than the recursive approach of Gel'fand and Pinsker for semideterministic broadcast channels. Based on our inner and outer bounds, we established the capacity regions for the semideterministic partially cooperative RBC, the orthogonal partially cooperative RBC, and the parallel partially cooperative RBC with unmatched degraded subchannels where $R_2 = 0$. We also established the capacity region for the parallel relay channel with unmatched degraded subchannels.

Our inner and outer bounds can be specialized to new bounds for the broadcast channel. By letting relay inputs be null, our inner bound reduces to a region that includes Marton's inner bound. Our outer bound reduces to a new capacity outer bound for the broadcast channel that generalizes Marton's outer bound to include a common message, and generalizes Gel'fand and Pinsker's outer bound to the general discrete memoryless broadcast channel.

## Appendix

## A Proof of Achievability for $\mathcal{R}'_1$ in Theorem 1

Suppose the source uses superposition encoding and binning, and destination 1 (the relay) uses the decode-and-forward scheme for the common message $W_0$. We adopt the regular encoding/sliding window decoding strategy [23] for the decode-and-forward scheme (see [6, Sec. I] for a review of relaying strategies).

We transmit in $B$ blocks that each have length $n$ (for convenience, we use the same notation for the block length as in Sec. 2 even though transmission is here done in smaller blocks). During each of the first $B-1$ blocks, a message tuple $(W_{0i}, W_{1i}, W_{2i}) \in [1, 2^{nR_0}] \times [1, 2^{nR_1}] \times [1, 2^{nR_2}]$ is encoded and sent from the source node, where $i$ denotes the index of the block and $i = 1, 2, \ldots, B-1$. The average rate triple $\left(R_0 \frac{B-1}{B}, R_1 \frac{B-1}{B}, R_2 \frac{B-1}{B}\right)$ over $B$ blocks approaches $(R_0, R_1, R_2)$ as $B \to \infty$.

We use random codes and fix a joint probability distribution

$$p(x_1)p(t|x_1)p(u_1, u_2|x_1, t)p(x|x_1, t, u_1, u_2)p(y_1, y_2|x, x_1). \quad (45)$$

Let $T_\epsilon^n(P_{X_1TU_1U_2XY_1Y_2})$ denote the strongly jointly $\epsilon$-typical set (see [19, Sec. 1.2]) based on the distribution (45).

*Random codebook generation:* We generate two statistically independent codebooks (codebooks 1 and 2) by following the steps outlined below twice. These two codebooks will be used for blocks with odd and even indices, respectively (see the *encoding* step).

1. Generate $2^{nR_0}$ independent identically distributed (i.i.d.) $x_1^n$ with distribution $\prod_{i=1}^n p(x_{1i})$. Index these codewords as $x_1^n(w_0')$, $w_0' \in [1, 2^{nR_0}]$.

2. For each $x_1^n(w_0')$, generate $2^{nR_0}$ i.i.d. $t^n$ with distribution $\prod_{i=1}^n p(t_i|x_{1i}(w_0'))$. Index these codewords as $t^n(w_0', w_0)$, $w_0 \in [1, 2^{nR_0}]$.

3. For each $x_1^n(w_0')$ and $t^n(w_0', w_0)$, generate $2^{n(R_1+R_1')}$ i.i.d. $u_1^n$ with distribution $\prod_{i=1}^n p(u_{1i}|t_i(w_0', w_0), x_{1i}(w_0'))$. Index these codewords as $u_1^n(w_0', w_0, w_1, v_1)$, $w_1 \in [1, 2^{nR_1}]$ and $v_1 \in [1, 2^{nR_1'}]$.

4. For each $x_1^n(w_0')$ and $t^n(w_0', w_0)$, generate $2^{n(R_2+R_2')}$ i.i.d. $u_2^n$ with distribution $\prod_{i=1}^n p(u_{2i}|t_i(w_0', w_0), x_{1i}(w_0'))$. Index these codewords as $u_2^n(w_0', w_0, w_2, v_2)$, $w_2 \in [1, 2^{nR_2}]$ and $v_2 \in [1, 2^{nR_2'}]$.

5. For each $x_1^n(w_0'), t^n(w_0', w_0), u_1^n(w_0', w_0, w_1, v_1)$ and $u_2^n(w_0', w_0, w_2, v_2)$, generate one $x^n$ with distribution

$$\prod_{i=1}^n p(x_i|u_{1i}(w_0', w_0, w_1, v_1), u_{2i}(w_0', w_0, w_2, v_2), t_i(w_0', w_0), x_{1i}(w_0')).$$

Denote $x^n$ by $x^n(w_0', w_0, w_1, v_1, w_2, v_2)$.

*Encoding:* We encode messages using codebooks 1 and 2, respectively, for blocks with odd and even indices. This is done because some of the decoding steps are performed jointly over two adjacent blocks, and having independent codebooks makes the error events corresponding to these blocks independent. The probabilities of these error events are thus easy to evaluate.

At the beginning of block $i$, let $(w_{0i}, w_{1i}, w_{2i})$ be the new message triple to be sent in block $i$, and $(w_{0[i-1]}, w_{1[i-1]}, w_{2[i-1]})$ be the message triple sent in block $i-1$. The source encoder first tries to find a pair $(v_{1i}, v_{2i})$ such that

$$\left(x_1^n(w_{0[i-1]}), t^n(w_{0[i-1]}, w_{0i}), u_1^n(w_{0[i-1]}, w_{0i}, w_{1i}, v_{1i}), u_2^n(w_{0[i-1]}, w_{0i}, w_{2i}, v_{2i})\right) \in T_\epsilon^n(P_{X_1TU_1U_2}).$$

One can show that such a pair exists with high probability for sufficiently large n if (see [16])

$$R_1' + R_2' > I(U_1; U_2|T, X_1). \quad (46)$$

The source then sends the codeword $x^n(w_{0[i-1]}, w_{0i}, w_{1i}, v_{1i}, w_{2i}, v_{2i})$.

At the beginning of block $i$, destination 1 (the relay node) has decoded the message $w_{0[i-1]}$ and transmits $x_1^n(w_{0[i-1]})$. For convenience, we list the codewords that are sent in the first three blocks in Table 1.

Table 1: Codewords being sent to achieve $\mathcal{R}_1'$ in Theorem 1

| block 1 | block 2 | block 3 |
|---|---|---|
| $x_1^n(1)$ | $x_1^n(w_{01})$ | $x_1^n(w_{02})$ |
| $t^n(1, w_{01})$ | $t^n(w_{01}, w_{02})$ | $t^n(w_{02}, w_{03})$ |
| $u_1^n(1, w_{01}, w_{11}, v_{11})$ | $u_1^n(w_{01}, w_{02}, w_{12}, v_{12})$ | $u_1^n(w_{02}, w_{03}, w_{13}, v_{13})$ |
| $u_2^n(1, w_{01}, w_{21}, v_{21})$ | $u_2^n(w_{01}, w_{02}, w_{22}, v_{22})$ | $u_2^n(w_{02}, w_{03}, w_{23}, v_{23})$ |

*Decoding:* The decoding procedures at the end of block $i$ are as follows:

1. Destination 1 knows $w_{0[i-1]}$ and declares the message pair $\left(\hat{w}_{0i}^{(1)}, \hat{w}_{1i}\right)$ is sent if there is a unique triple $\left(\hat{w}_{0i}^{(1)}, \hat{w}_{1i}, \hat{v}_{1i}\right)$ such that

$$\left(x_1^n(w_{0[i-1]}), t^n\left(w_{0[i-1]}, \hat{w}_{0i}^{(1)}\right), u_1^n\left(w_{0[i-1]}, \hat{w}_{0i}^{(1)}, \hat{w}_{1i}, \hat{v}_{1i}\right), y_{1i}^n\right) \in T_\epsilon^n(P_{X_1 T U_1 Y_1}).$$

One can show that the decoding error in this step is small for sufficiently large $n$ if

$$\begin{aligned} R_1 + R_1' &< I(U_1; Y_1 | T, X_1) \\ R_0 + R_1 + R_1' &< I(T, U_1; Y_1 | X_1). \end{aligned} \quad (47)$$

2. Destination 2 knows $w_{0[i-2]}$ and decodes $(w_{0[i-1]}, w_{2[i-1]})$ based on the information received in block $i-1$ and block $i$. It declares that the message pair $\left(\hat{w}_{0[i-1]}^{(2)}, \hat{w}_{2[i-1]}\right)$ is sent if there is a unique tuple $\left(\hat{w}_{0[i-1]}^{(2)}, \hat{w}_{2[i-1]}, \hat{v}_{2[i-1]}\right)$ such that

$$\left(x_1^n(w_{0[i-2]}), t^n\left(w_{0[i-2]}, \hat{w}_{0[i-1]}^{(2)}\right), u_2^n\left(w_{0[i-2]}, \hat{w}_{0[i-1]}^{(2)}, \hat{w}_{2[i-1]}, \hat{v}_{2[i-1]}\right), y_{2[i-1]}^n\right) \in T_\epsilon^n(P_{X_1 T U_2 Y_2}),$$

and $\quad \left(x_1^n\left(\hat{w}_{0[i-1]}^{(2)}\right), y_{2i}^n\right) \in T_\epsilon^n(P_{X_1 Y_2}).$

One can show that the decoding error in this step is small for sufficiently large $n$ if

$$\begin{aligned} R_2 + R_2' &< I(U_2; Y_2 | T, X_1) \\ R_0 + R_2 + R_2' &< I(T, X_1, U_2; Y_2). \end{aligned} \quad (48)$$

Combining (46), (47), and (48), we find that $\mathcal{R}_1'$ is achievable.

# B  Proof of Theorem 4

Consider a code with length $n$ and average block error probability $P_e$. The probability distribution on the joint ensemble space $W_0 \times W_1 \times W_2 \times \mathcal{X}^n \times \mathcal{X}_1^n \times \mathcal{Y}_1^n \times \mathcal{Y}_2^n$ is given by

$$p(w_0, w_1, w_2, x^n, x_1^n, y_1^n, y_2^n)$$
$$= p(w_0)p(w_1)p(w_2)p(x^n|w_0, w_1, w_2) \prod_{i=1}^{n} \left[ f_i(x_{1i}|y_1^{i-1}) p(y_{1i}, y_{2i}|x_i, x_{1i}) \right]. \tag{49}$$

By Fano's inequality [24, Sec. 2.11], we have

$$\begin{aligned} H(W_0|Y_1^n) &\leq H(W_0, W_1|Y_1^n) \leq n(R_0 + R_1)P_e + 1 := n\delta_1, \\ H(W_0|Y_2^n) &\leq H(W_0, W_2|Y_2^n) \leq n(R_0 + R_2)P_e + 1 := n\delta_2 \end{aligned} \tag{50}$$

where $\delta_1, \delta_2 \to 0$ if $P_e \to 0$. Similarly, we have

$$\begin{aligned} H(W_1|Y_1^n) &\leq H(W_0, W_1|Y_1^n) \leq n\delta_1, \\ H(W_2|Y_2^n) &\leq H(W_0, W_2|Y_2^n) \leq n\delta_2. \end{aligned} \tag{51}$$

We define the following auxiliary random variables:

$$T_i := (W_0, Y_1^{i-1}, Y_{2[i+1]}^n), \quad \text{for } i = 1, 2, \ldots, n; \tag{52}$$
$$U_i := (W_2, Y_1^{i-1}, Y_{2[i+1]}^n), \quad \text{for } i = 1, 2, \ldots, n. \tag{53}$$

We first bound the common rate $R_0$:

$$\begin{aligned} nR_0 &= H(W_0) = I(W_0; Y_1^n) + H(W_0|Y_1^n) \\ &\stackrel{(a)}{\leq} I(W_0; Y_1^n) + n\delta_1 \\ &= \sum_{i=1}^{n} I(W_0; Y_{1i}|Y_1^{i-1}) + n\delta_1 \\ &\stackrel{(b)}{=} \sum_{i=1}^{n} I(W_0; Y_{1i}|Y_1^{i-1}, X_{1i}) + n\delta_1 \\ &\stackrel{(c)}{\leq} \sum_{i=1}^{n} H(Y_{1i}|X_{1i}) - H(Y_{1i}|W_0, Y_1^{i-1}, X_{1i}, Y_{2[i+1]}^n) + n\delta_1 \\ &\stackrel{(d)}{=} \sum_{i=1}^{n} H(Y_{1i}|X_{1i}) - H(Y_{1i}|T_i, X_{1i}) + n\delta_1 \\ &= \sum_{i=1}^{n} I(T_i; Y_{1i}|X_{1i}) + n\delta_1 \end{aligned} \tag{54}$$

where (a) follows from Fano's inequality in (50), (b) follows because $X_{1i}$ is a function of $Y_1^{i-1}$, (c) follows because conditioning does not increase entropy, and (d) follows from the definition of $T_i$ given in (52).

We can similarly derive the following upper bound on $R_0$:

$$\begin{aligned}
nR_0 &= H(W_0) = I(W_0; Y_2^n) + H(W_0|Y_2^n) \\
&\leq I(W_0; Y_2^n) + n\delta_2 \\
&= \sum_{i=1}^{n} I(W_0; Y_{2i}|Y_{2[i+1]}^n) + n\delta_2 \\
&\stackrel{(a)}{\leq} \sum_{i=1}^{n} H(Y_{2i}|Y_{2[i+1]}^n) - H(Y_{2i}|W_0, Y_1^{i-1}, Y_{2[i+1]}^n, X_{1i}) + n\delta_2 \qquad (55) \\
&\leq \sum_{i=1}^{n} H(Y_{2i}) - H(Y_{2i}|T_i, X_{1i}) + n\delta_2 \\
&= \sum_{i=1}^{n} I(T_i, X_{1i}; Y_{2i}) + n\delta_2.
\end{aligned}$$

where $(a)$ follows because conditioning does not increase entropy. We now bound the sum rate $R_0 + R_1$:

$$\begin{aligned}
nR_0 + nR_1 &= H(W_0, W_1) = I(W_0, W_1; Y_1^n) + H(W_0, W_1|Y_1^n) \\
&\leq I(W_0, W_1; Y_1^n) + n\delta_1 \\
&= \sum_{i=1}^{n} I(W_0, W_1; Y_{1i}|Y_1^{i-1}) + n\delta_1 \\
&\stackrel{(a)}{=} \sum_{i=1}^{n} I(W_0, W_1; Y_{1i}|Y_1^{i-1}, X_{1i}) + n\delta_1 \\
&\leq \sum_{i=1}^{n} H(Y_{1i}|Y_1^{i-1}, X_{1i}) - H(Y_{1i}|Y_1^{i-1}, W_0, W_1, X_{1i}, X_i) + n\delta_1 \qquad (56) \\
&\stackrel{(b)}{\leq} \sum_{i=1}^{n} H(Y_{1i}|X_{1i}) - H(Y_{1i}|X_{1i}, X_i) + n\delta_1 \\
&= \sum_{i=1}^{n} I(X_i; Y_{1i}|X_{1i}) + n\delta_1.
\end{aligned}$$

where $(a)$ follows because $X_{1i}$ is a function of $Y_1^{i-1}$, and $(b)$ follows from the Markov chain condition $(Y_1^{i-1}, W_0, W_1) \to (X_{1i}, X_i) \to Y_{1i}$.

We next derive an upper bound on the sum rate $R_0 + R_2$:

$$\begin{aligned}
nR_0 + nR_2 &= H(W_0, W_2) = I(W_0, W_2; Y_2^n) + H(W_0, W_2|Y_2^n) \\
&\leq I(W_0, W_2; Y_2^n) + n\delta_2 \\
&= \sum_{i=1}^{n} I(W_0, W_2; Y_{2i}|Y_{2[i+1]}^n) + n\delta_2 \\
&\stackrel{(a)}{\leq} \sum_{i=1}^{n} H(Y_{2i}) - H(Y_{2i}|W_0, W_2, Y_{2[i+1]}^n, Y_1^{i-1}, X_{1i}) + n\delta_2 \qquad (57) \\
&\stackrel{(b)}{\leq} \sum_{i=1}^{n} H(Y_{2i}) - H(Y_{2i}|T_i, U_i, X_{1i}) + n\delta_2 \\
&= \sum_{i=1}^{n} I(T_i, U_i, X_{1i}; Y_{2i}) + n\delta_2
\end{aligned}$$

where $(a)$ follows because conditioning does not increase entropy, and $(b)$ follows from the definitions of $T_i$ and $U_i$ given in (52) and (53), respectively. We also have the following upper bound on the sum rate $R_0 + R_2$:

$$\begin{aligned}
nR_0 + nR_2 &= H(W_0, W_2) = I(W_0, W_2; Y_2^n) + H(W_0, W_2|Y_2^n) \\
&\leq I(W_0, W_2; Y_2^n) + n\delta_2 \\
&= \sum_{i=1}^{n} I(W_0, W_2; Y_{2i}|Y_2^{i-1}) + n\delta_2 \\
&\leq \sum_{i=1}^{n} H(Y_{2i}) - H(Y_{2i}|W_0, W_2, Y_2^{i-1}, X_{1i}, X_i) + n\delta_2 \qquad (58) \\
&\stackrel{(a)}{\leq} \sum_{i=1}^{n} H(Y_{2i}) - H(Y_{2i}|X_i, X_{1i}) + n\delta_2 \\
&= \sum_{i=1}^{n} I(X_i, X_{1i}; Y_{2i}) + n\delta_2
\end{aligned}$$

where $(a)$ follows from the Markov chain condition $(W_0, W_2, Y_2^{i-1}) \to (X_{1i}, X_i) \to Y_{2i}$.

We now introduce the following lemmas that will be useful for bounding the sum rate $R_0 + R_1 + R_2$.

**Lemma 2.**

$$\sum_{i=1}^{n} I(Y_{2i}; Y_1^{i-1}|W_0, W_2, Y_{2[i+1]}^n) = \sum_{i=1}^{n} I(Y_{1i}; Y_{2[i+1]}^n|W_0, W_2, Y_1^{i-1}),$$
$$\sum_{i=1}^{n} I(Y_{2i}; Y_1^{i-1}|W_0, Y_{2[i+1]}^n) = \sum_{i=1}^{n} I(Y_{1i}; Y_{2[i+1]}^n|W_0, Y_1^{i-1}).$$

*Proof.* See [25, Lemma 7]. □

**Lemma 3.**

$$I(W_0, W_2; Y_2^n) = \sum_{i=1}^{n} I(W_0, W_2, Y_1^{i-1}; Y_{2i}^n) - I(W_0, W_2, Y_1^i; Y_{2[i+1]}^n) \quad (59)$$

where $Y_{10} = 0$.

*Proof.* By direct calculation. □

Define $\delta := 2\delta_1 + \delta_2$. We obtain the following bound on $R_0 + R_1 + R_2$:

$$\begin{aligned}
nR_0 &+ nR_1 + nR_2 \\
&\leq I(W_0; Y_1^n) + I(W_1; Y_1^n) + I(W_2; Y_2^n) + n\delta \\
&\leq I(W_0; Y_1^n) + I(W_1; Y_1^n, W_0, W_2) + I(W_2; Y_2^n, W_0) + n\delta \\
&\stackrel{(a)}{=} I(W_0; Y_1^n) + I(W_1; Y_1^n | W_0, W_2) + I(W_2; Y_2^n | W_0) \\
&\quad + I(W_2; Y_1^n | W_0) - I(W_2; Y_1^n | W_0) + n\delta \\
&= I(W_0, W_1, W_2; Y_1^n) - I(W_2; Y_1^n | W_0) + I(W_2; Y_2^n | W_0) + n\delta
\end{aligned} \quad (60)$$

where $(a)$ follows because $W_0$, $W_1$ and $W_2$ are independent. For the first term in (60), we obtain the following bound:

$$\begin{aligned}
I(W_0, &W_1, W_2; Y_1^n) \\
&\stackrel{(a)}{=} \sum_{i=1}^{n} I(W_0, W_1, W_2; Y_{1i} | Y_1^{i-1}, X_{1i}) \\
&\leq \sum_{i=1}^{n} H(Y_{1i} | X_{1i}) - H(Y_{1i} | W_0, W_1, W_2, Y_1^{i-1}, X_{1i}, X_i) \\
&\stackrel{(b)}{=} \sum_{i=1}^{n} H(Y_{1i} | X_{1i}) - H(Y_{1i} | W_0, Y_1^{i-1}, X_{1i}, X_i) \\
&\stackrel{(c)}{\leq} \sum_{i=1}^{n} H(Y_{1i} | X_{1i}) - H(Y_{1i} | W_0, Y_1^{i-1}, Y_{2[i+1]}^n, X_{1i}, X_i) \\
&= \sum_{i=1}^{n} H(Y_{1i} | X_{1i}) - H(Y_{1i} | T_i, X_{1i}, X_i) \\
&= \sum_{i=1}^{n} I(T_i, X_i; Y_{1i} | X_{1i})
\end{aligned} \quad (61)$$

where $(a)$ follows from the chain rule and because $X_{1i}$ is a function of $Y_1^{i-1}$, $(b)$ follows from the Markov chain condition: $(W_0, W_1, W_2, Y_1^{i-1}) \to (X_{1i}, X_i) \to Y_{1i}$, and $(c)$ follows because conditioning does not increase entropy. For the sum of the second and third terms in (60), we obtain the following bound:

$$
\begin{aligned}
-I(W_2; & Y_1^n|W_0) + I(W_2; Y_2^n|W_0) \\
&= \sum_{i=1}^{n} -I(W_2; Y_{1i}|W_0, Y_1^{i-1}) + I(W_2; Y_{2i}|W_0, Y_{2[i+1]}^n) \\
&\stackrel{(a)}{=} \sum_{i=1}^{n} -I(W_2, Y_{2[i+1]}^n; Y_{1i}|W_0, Y_1^{i-1}) + I(Y_{2[i+1]}^n; Y_{1i}|W_0, W_2, Y_1^{i-1}) \\
&\qquad + I(W_2, Y_1^{i-1}; Y_{2i}|W_0, Y_{2[i+1]}^n) - I(Y_1^{i-1}; Y_{2i}|W_0, W_2, Y_{2[i+1]}^n) \\
&\stackrel{(b)}{=} \sum_{i=1}^{n} -I(W_2, Y_{2[i+1]}^n; Y_{1i}|W_0, Y_1^{i-1}) + I(W_2, Y_1^{i-1}; Y_{2i}|W_0, Y_{2[i+1]}^n) \\
&\stackrel{(c)}{=} \sum_{i=1}^{n} -I(Y_{2[i+1]}^n; Y_{1i}|W_0, Y_1^{i-1}) - I(W_2; Y_{1i}|W_0, Y_1^{i-1}, Y_{2[i+1]}^n, X_{1i}) \\
&\qquad + I(Y_1^{i-1}; Y_{2i}|W_0, Y_{2[i+1]}^n) + I(W_2; Y_{2i}|W_0, Y_{2[i+1]}^n, Y_1^{i-1}, X_{1i}) \\
&\stackrel{(d)}{=} \sum_{i=1}^{n} -I(W_2; Y_{1i}|W_0, Y_1^{i-1}, Y_{2[i+1]}^n, X_{1i}) + I(W_2; Y_{2i}|W_0, Y_{2[i+1]}^n, Y_1^{i-1}, X_{1i}) \\
&\stackrel{(e)}{=} \sum_{i=1}^{n} -I(U_i; Y_{1i}|T_i, X_{1i}) + I(U_i; Y_{2i}|T_i, X_{1i})
\end{aligned}
\tag{62}
$$

where $(a)$ follows from the chain rule, $(b)$ follows from Lemma 2, $(c)$ follows from the chain rule and the fact that $X_{1i}$ is a function of $Y_1^{i-1}$, $(d)$ follows from Lemma 2, and $(e)$ follows from the definitions of $T_i$ and $U_i$ given in (52) and (53), respectively.

We substitute the bounds (61) and (62) into (60) and obtain:

$$\begin{aligned}
nR_0 &+ nR_1 + nR_2 \\
&\leq \sum_{i=1}^{n} I(T_i, X_i; Y_{1i}|X_{1i}) - I(U_i; Y_{1i}|T_i, X_{1i}) + I(U_i; Y_{2i}|T_i, X_{1i}) + n\delta \\
&= \sum_{i=1}^{n} I(T_i; Y_{1i}|X_{1i}) + I(X_i; Y_{1i}|T_i, X_{1i}) - I(U_i; Y_{1i}|T_i, X_{1i}) \\
&\qquad + I(U_i; Y_{2i}|T_i, X_{1i}) + n\delta \\
&= \sum_{i=1}^{n} I(T_i; Y_{1i}|X_{1i}) + H(Y_{1i}|T_i, X_{1i}) - H(Y_{1i}|T_i, X_{1i}, X_i) \\
&\qquad - H(Y_{1i}|T_i, X_{1i}) + H(Y_{1i}|T_i, X_{1i}, U_i) + I(U_i; Y_{2i}|T_i, X_{1i}) \qquad (63) \\
&= \sum_{i=1}^{n} I(T_i; Y_{1i}|X_{1i}) - H(Y_{1i}|T_i, X_{1i}, X_i) + H(Y_{1i}|T_i, X_{1i}, U_i) \\
&\qquad + I(U_i; Y_{2i}|T_i, X_{1i}) \\
&\stackrel{(a)}{\leq} \sum_{i=1}^{n} I(T_i; Y_{1i}|X_{1i}) - H(Y_{1i}|T_i, X_{1i}, X_i, U_i) + H(Y_{1i}|T_i, X_{1i}, U_i) \\
&\qquad + I(U_i; Y_{2i}|T_i, X_{1i}) + n\delta \\
&= \sum_{i=1}^{n} I(T_i; Y_{1i}|X_{1i}) + I(X_i; Y_{1i}|T_i, X_{1i}, U_i) + I(U_i; Y_{2i}|T_i, X_{1i}) + n\delta
\end{aligned}$$

where $(a)$ follows because conditioning does not increase entropy.

We can also bound the sum rate $R_0 + R_1 + R_2$ as follows:

$$\begin{aligned}
nR_0 &+ nR_1 + nR_2 \\
&\leq I(W_1; Y_1^n) + I(W_0, W_2; Y_2^n) + n\delta \\
&\leq I(W_1; Y_1^n, W_0, W_2) + I(W_0, W_2; Y_2^n) + n\delta \\
&= I(W_1; Y_1^n | W_0, W_2) + I(W_0, W_2; Y_2^n) + n\delta \\
&\stackrel{(a)}{=} \sum_{i=1}^n I(W_1; Y_{1i} | W_0, W_2, Y_1^{i-1}) + I(W_0, W_2, Y_1^{i-1}; Y_{2i}^n) - I(W_0, W_2, Y_1^i; Y_{2[i+1]}^n) + n\delta \\
&\stackrel{(b)}{=} \sum_{i=1}^n I(W_1; Y_{1i} | W_0, W_2, Y_1^{i-1}) + I(W_0, W_2, Y_1^{i-1}; Y_{2[i+1]}^n) + I(W_0, W_2, Y_1^{i-1}; Y_{2i} | Y_{2[i+1]}^n) \\
&\quad - I(W_0, W_2, Y_1^{i-1}; Y_{2[i+1]}^n) - I(Y_{1i}; Y_{2[i+1]}^n | W_0, W_2, Y_1^{i-1}) + n\delta \\
&= \sum_{i=1}^n I(W_1; Y_{1i} | W_0, W_2, Y_1^{i-1}) + I(W_0, W_2, Y_1^{i-1}; Y_{2i} | Y_{2[i+1]}^n) \\
&\quad - I(Y_{1i}; Y_{2[i+1]}^n | W_0, W_2, Y_1^{i-1}) + n\delta \\
&= \sum_{i=1}^n -H(Y_{1i} | W_0, W_2, W_1, Y_1^{i-1}) + I(W_0, W_2, Y_1^{i-1}; Y_{2i} | Y_{2[i+1]}^n) \\
&\quad + H(Y_{1i} | W_0, W_2, Y_1^{i-1}, Y_{2[i+1]}^n) + n\delta \\
&\stackrel{(c)}{\leq} \sum_{i=1}^n -H(Y_{1i} | W_0, W_2, W_1, Y_1^{i-1}, X_{1i}, X_i) + I(W_0, W_2, Y_1^{i-1}, Y_{2[i+1]}^n, X_{1i}; Y_{2i}) \\
&\quad + H(Y_{1i} | W_0, W_2, Y_1^{i-1}, Y_{2[i+1]}^n, X_{1i}) + n\delta \\
&\stackrel{(d)}{=} \sum_{i=1}^n -H(Y_{1i} | W_0, W_2, Y_1^{i-1}, X_{1i}, X_i) + I(T_i, U_i, X_{1i}; Y_{2i}) \\
&\quad + H(Y_{1i} | T_i, U_i, X_{1i}) + n\delta \\
&\stackrel{(e)}{\leq} \sum_{i=1}^n -H(Y_{1i} | W_0, W_2, Y_1^{i-1}, Y_{2[i+1]}^n, X_{1i}, X_i) + I(T_i, U_i, X_{1i}; Y_{2i}) \\
&\quad + H(Y_{1i} | T_i, U_i, X_{1i}) + n\delta \\
&= \sum_{i=1}^n I(X_i; Y_{1i} | T_i, U_i, X_{1i}) + I(T_i, U_i, X_{1i}; Y_{2i}) + n\delta
\end{aligned}$$
(64)

where $(a)$ follows from Lemma 3, $(b)$ follows from the chain rule, $(c)$ follows because conditioning does not increase entropy, because $I(A; B|C) \leq I(A, C; B)$, and because $X_{1i}$ is a function of $Y_1^{i-1}$, $(d)$ follows from the Markov chain condition $(W_0, W_2, W_1, Y_1^{i-1}) \to (X_{1i}, X_i) \to Y_{1i}$, and from the definitions of $T_i$ and $U_i$, and $(e)$ follows because conditioning does not increase entropy.

Finally, consider the following bound on $R_0 + R_1 + R_2$:

$$\begin{aligned}
nR_0 + nR_1 + nR_2 &\leq I(W_0, W_1, W_2; Y_1^n, Y_2^n) + n\delta \\
&= \sum_{i=1}^{n} I(W_0, W_1, W_2; Y_{1i}, Y_{2i} | Y_1^{i-1}, Y_2^{i-1}) + n\delta \\
&\stackrel{(a)}{=} \sum_{i=1}^{n} I(W_0, W_1, W_2; Y_{1i}, Y_{2i} | Y_1^{i-1}, Y_2^{i-1}, X_{1i}) + n\delta \\
&\stackrel{(b)}{\leq} \sum_{i=1}^{n} H(Y_{1i}, Y_{2i} | Y_1^{i-1}, Y_2^{i-1}, X_{1i}) \\
&\quad - H(Y_{1i}, Y_{2i} | W_0, W_1, W_2, Y_1^{i-1}, Y_2^{i-1}, X_{1i}, X_i) + n\delta \\
&\stackrel{(c)}{\leq} \sum_{i=1}^{n} H(Y_{1i}, Y_{2i} | X_{1i}) - H(Y_{1i}, Y_{2i} | X_{1i}, X_i) + n\delta \\
&\leq \sum_{i=1}^{n} I(X_i; Y_{1i}, Y_{2i} | X_{1i}) + n\delta.
\end{aligned} \quad (65)$$

where $(a)$ follows because $X_{1i}$ is a function of $Y_1^{i-1}$, $(b)$ follows because conditioning does not increase entropy, and $(c)$ follows from the Markov chain condition:

$$(W_0, W_1, W_2, Y_1^{i-1}, Y_2^{i-1}) \to (X_{1i}, X_i) \to (Y_{1i}, Y_{2i}). \quad (66)$$

Note that the random variables $T^n, U^n, X^n, X_1^n, Y_1^n, Y_2^n$ do not satisfy the Markov chain condition: $(T_i, U_i) \to (X_{1i}, X_i) \to (Y_{1i}, Y_{2i})$. This is because at time $i$, the current output $Y_{1i}$ affects the future relay inputs $X_{1[i+1]}^n$, which affect the future outputs $Y_{2[i+1]}^n$ that are contained in random variables $T_i$ and $U_i$. Hence $Y_{1i}$ can be correlated with $T_i$ and $U_i$ even conditioned on the current inputs $X_{1i}$ and $X_i$. However, the Markov chain condition $(T_i, U_i) \to (X_{1i}, X_i, Y_{1i}) \to Y_{2i}$ is satisfied.

We now reduce the upper bounds to a single letter characterization. We introduce a random variable $Q$ that is independent of $W_0, W_1, W_2, X^n, X_1^n, Y_1^n, Y_2^n$ and is uniformly distributed over $\{1, 2, \ldots, n\}$. Define $T = (Q, T_Q)$, $U = (Q, U_Q)$, $X = X_Q$, $X_1 = X_{1Q}$, $Y_1 = Y_{1Q}$, and $Y_2 = Y_{2Q}$. By using these definitions, the bounds (54)-(58) and (63)-(65) can be transformed into the single letter bounds in Theorem 4 by following standard steps (e.g., [3, Sec. III]).

## C  Proof of Achievability and Geometric Illustration for $\mathcal{R}_3$ in Theorem 3

Although the achievability of $\mathcal{R}_3$ can be directly seen from the achievability of $\mathcal{R}_2$ in Theorem 2, it is instructive to derive $\mathcal{R}_3$ from $\mathcal{R}_1'$ in Theorem 1 by exploring the geometric structure of $\mathcal{R}_3$. In the following proof, we assume that $I(T; Y_1|X_1) \leq I(T, X_1; Y_2)$. The case where $I(T; Y_1|X_1) > I(T, X_1; Y_2)$ can be proved in a similar way as outlined below.

The region $\mathcal{R}_3$ in Theorem 3 can have five possible structures (see Fig. 10) depending on how the bounds on the rates $R_0$, $R_0 + R_1$, $R_0 + R_2$, and $R_0 + R_1 + R_2$ compare with each other. We will show that the region corresponding to case 5 is achievable, where the bound on the sum rate $R_0 + R_1 + R_2$ is larger than the bounds on $R_0$, $R_0 + R_1$ and $R_0 + R_2$. The achievability of the regions for the other four cases is either easy (binning is not necessary) or can be similarly proved.

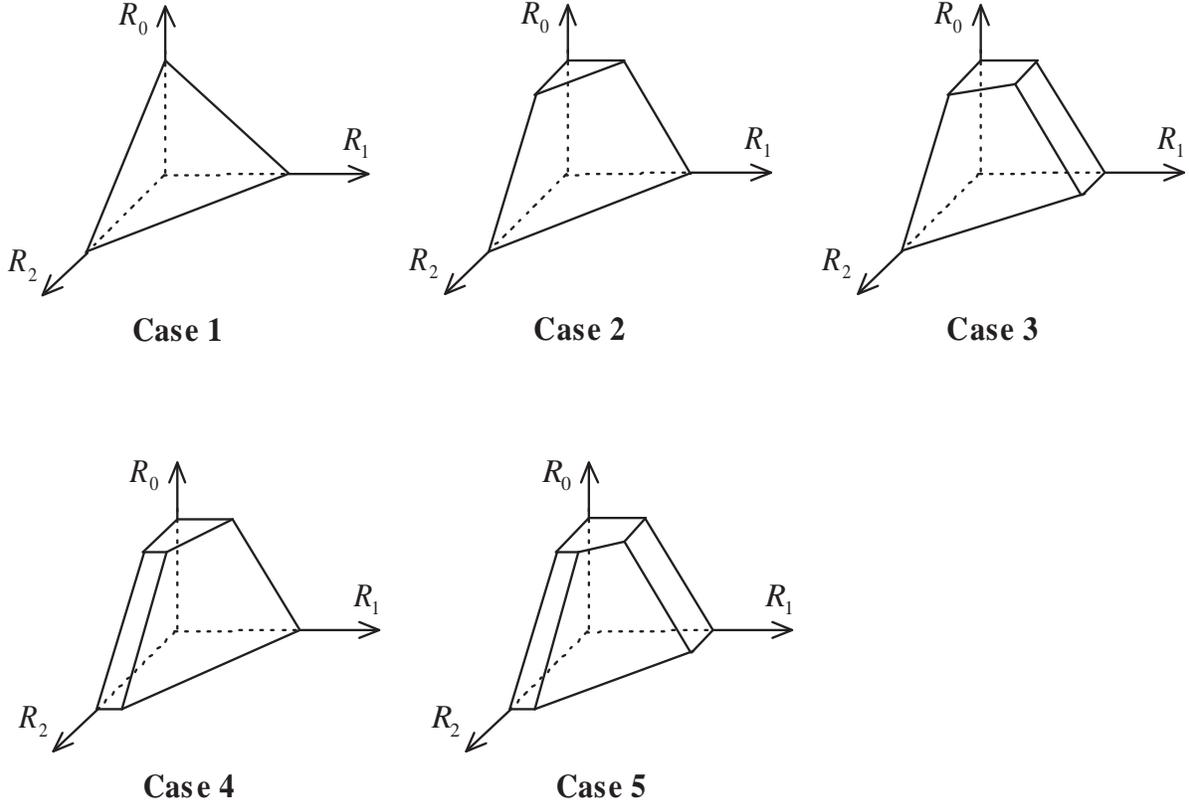

Figure 10: Five cases of the region $\mathcal{R}_3$

We plot $\mathcal{R}_3$ corresponding to case 5 in more detail in Fig. 11. To prove this region is achievable, it suffices to show that the two corner points A and B are achievable. The rate triples in the rest of the region can then be achieved either by time-sharing or by switching part of $R_0$ to $R_1$ or $R_2$ according to Lemma 1.

The rate triple for corner point A is given by

$$\begin{aligned} R_0^A &= I(T; Y_1|X_1), \\ R_1^A &= I(U_1; Y_1|T, X_1), \\ R_2^A &= I(U_2; Y_2|T, X_1) - I(U_1; U_2|T, X_1). \end{aligned} \quad (67)$$

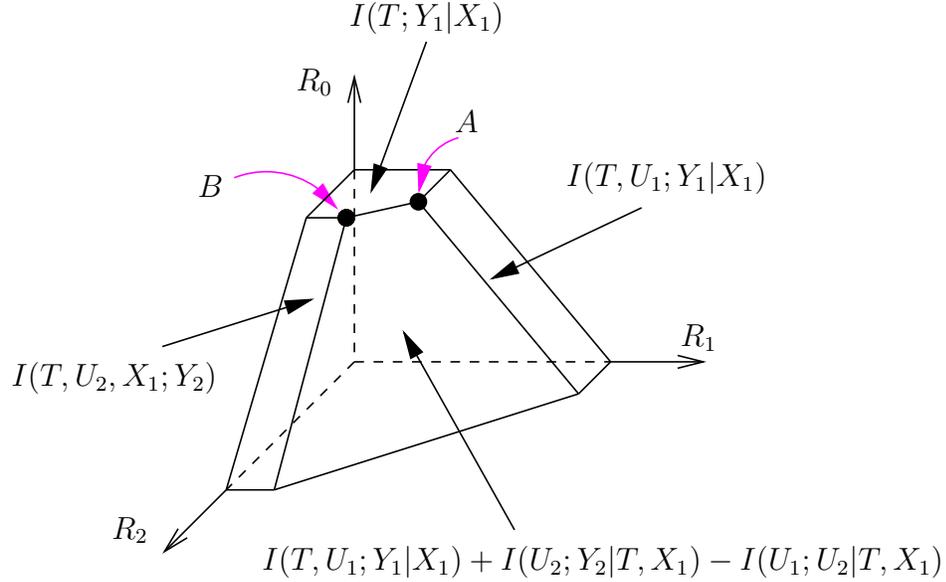

Figure 11: Case 5 of the region $\mathcal{R}_3$

It is easy to check that this point is in $\mathcal{R}'_1$ in Theorem 1 by choosing $R'_1 = 0$ and $R'_2 = I(U_1; U_2 | T, X_1)$. Hence the corner point A is achievable.

The rate triple for corner point B is given by

$$
\begin{aligned}
R_0^B &= I(T; Y_1 | X_1), \\
R_1^B &= I(T, U_1; Y_1 | X_1) - I(T, X_1; Y_2) - I(U_1; U_2 | T, X_1), \\
R_2^B &= I(T, U_2, X_1; Y_2) - I(T; Y_1 | X_1).
\end{aligned} \quad (68)
$$

One can also check that the following rate triple is in $\mathcal{R}'_1$ by choosing $R'_1 = I(U_1; U_2 | T, X_1)$ and $R'_2 = 0$:

$$
\begin{aligned}
R_0 &= I(T, X_1; Y_2), \\
R_1 &= I(T, U_1; Y_1 | X_1) - I(T, X_1; Y_2) - I(U_1; U_2 | T, X_1), \\
R_2 &= I(U_2; Y_2 | T, X_1).
\end{aligned} \quad (69)
$$

But the rate triple in (68) can be expressed in terms of (69) as $(R_0^B, R_1^B, R_2^B) = (R_0 - \delta, R_1, R_2 + \delta)$ where $\delta = I(T, X_1; Y_2) - I(T; Y_1 | X_1)$. Hence the corner point B is achievable by Lemma 1 in Section 3.

# D  Proof of Achievability for Theorem 9

Suppose destination 1 (the relay) uses the decode-and-forward scheme by forwarding only $W_0$. We transmit in $B$ blocks that each have length $n$ (we again use the same notation for the block length as in Section 2). For the first $B - 1$ blocks, a message pair $(W_{0i}, W_{1i}) \in$ $[1, 2^{nR_0}] \times [1, 2^{nR_1}]$ is encoded into a codeword $x_R^n$ and sent through the source-to-relay channel, where $i$ denotes the index of the block and $i = 1, 2, \ldots, B - 1$. For the blocks indexed by $i = 2, \ldots, B$, a message pair $(W_{0[i-1]}, W_{2[i-1]}) \in [1, 2^{nR_0}] \times [1, 2^{nR_2}]$ is encoded into a codeword $x_D^n$ and sent through the channel from the source to destination 2. The average rate triple $(R_0 \frac{B-1}{B}, R_1 \frac{B-1}{B}, R_2 \frac{B-1}{B})$ over $B$ blocks approaches $(R_0, R_1, R_2)$ as $B \to \infty$.

We use random codes and fix a probability distribution

$$p(x_1)p(x_R|x_1)p(x_D|x_1)p(y_1|x_R, x_1)p(y_2|x_D, x_2). \tag{70}$$

We use $T_\epsilon^n(P_{X_1 X_R X_D Y_1 Y_2})$ to denote the jointly $\epsilon$-typical set based on the distribution (70).

*Random codebook generation:* We generate a codebook as follows:

1. Generate $2^{nR_0}$ i.i.d. $x_1^n$ with distribution $\prod_{i=1}^n p(x_{1i})$. Index these codewords as $x_1^n(w_0')$, $w_0' \in [1, 2^{nR_0}]$.

2. For each $x_1^n(w_0')$, generate $2^{n(R_0+R_1)}$ i.i.d. $x_R^n$ with distribution $\prod_{i=1}^n p(x_{Ri}|x_{1i}(w_0'))$. Index these codewords as $x_R^n(w_0', w_0, w_1)$, $w_0 \in [1, 2^{nR_0}]$ and $w_1 \in [1, 2^{nR_1}]$.

3. For each $x_1^n(w_0')$, generate $2^{nR_2}$ i.i.d. $x_D^n$ with distribution $\prod_{i=1}^n p(x_{Di}|x_{1i}(w_0'))$. Index these codewords as $x_D^n(w_0', w_2)$, $w_2 \in [1, 2^{nR_2}]$.

*Encoding:* At the beginning of block $i$, the source sends $x_R^n(w_{0[i-1]}, w_{0i}, w_{1i})$ and $x_D^n(w_{0[i-1]}, w_{2[i-1]})$. The relay (destination 1) has decoded $w_{0[i-1]}$ transmitted in the previous block. It then sends the codeword $x_1^n(w_{0[i-1]})$. For convenience, we list the codewords that are sent in the first three blocks in Table 2.

Table 2: Codewords being sent to achieve $\mathcal{C}_{or}$ in Theorem 9

| block 1 | block 2 | block 3 |
| --- | --- | --- |
| $x_1^n(1)$ | $x_1^n(w_{01})$ | $x_1^n(w_{02})$ |
| $x_R^n(1, w_{01}, w_{11})$ | $x_R^n(w_{01}, w_{02}, w_{12})$ | $x_R^n(w_{02}, w_{03}, w_{13})$ |
| $x_D^n(1, 1)$ | $x_D^n(w_{01}, w_{21})$ | $x_D^n(w_{02}, w_{22})$ |

*Decoding:* The decoding procedures at the end of block $i$ are as follows:

1. Destination 1 knows $w_{0[i-1]}$ and declares the message pair $(\hat{w}_{0i}^{(1)}, \hat{w}_{1i})$ is sent if there is a unique pair $(\hat{w}_{0i}^{(1)}, \hat{w}_{1i})$ such that

$$\left( x_1^n(w_{0[i-1]}), x_R^n(w_{0[i-1]}, \hat{w}_{0i}^{(1)}, \hat{w}_{1i}), y_{1i}^n \right) \in T_\epsilon^n(P_{X_1 X_R Y_1}).$$

One can show that the decoding error in this step is small for sufficiently large $n$ if

$$R_0 + R_1 < I(X_R; Y_1 | X_1). \tag{71}$$

2. Destination 2 declares that the message pair $(\hat{w}_{0[i-1]}^{(2)}, \hat{w}_{2[i-1]})$ is sent if there is a unique pair $(\hat{w}_{0[i-1]}^{(2)}, \hat{w}_{2[i-1]})$ such that

$$\left(x_1^n(\hat{w}_{0[i-1]}^{(2)}), x_D^n(\hat{w}_{0[i-1]}^{(2)}, \hat{w}_{2[i-1]}), y_{2i}^n\right) \in T_\epsilon^n(P_{X_1 X_D Y_2}).$$

One can show that the decoding error in this step is small for sufficiently large $n$ if

$$\begin{aligned} R_2 &< I(X_D; Y_2|X_1) \\ R_0 + R_2 &< I(X_D, X_1; Y_2). \end{aligned} \quad (72)$$

Combining (71) and (72), we find that the following region is achievable:

$$\mathcal{R}'_{or} = \bigcup_{\substack{p(x_1)p(x_R|x_1)p(x_D|x_1) \\ p(y_1|x_R, x_1)p(y_2|x_D, x_1)}} \left\{ \begin{array}{l} (R_0, R_1, R_2): \\ R_0 \geq 0, R_1 > 0, R_2 > 0, \\ R_0 + R_1 < I(X_R; Y_1|X_1), \\ R_0 + R_2 < I(X_D, X_1; Y_2), \\ R_2 < I(X_D; Y_2|X_1). \end{array} \right\} \quad (73)$$

We next show that the achievability of $\mathcal{R}'_{or}$ implies the achievability of $\mathcal{C}_{or}$ in Theorem 9. Suppose that $(R'_0, R'_1, R'_2) \in \mathcal{C}_{or}$; if $R'_2 < I(X_D; Y_2|X_1)$ then $(R'_0, R'_1, R'_2) \in \mathcal{R}'_{or}$ and this rate triple is hence achievable. If $R'_2 \geq I(X_D; Y_2|X_1)$, let $R'_2 = I(X_D; Y_2|X_1) + \alpha$. Note that $0 \leq \alpha < I(X_R; Y_1|X_1)$ from (28). Consider

$$R'_0 + R'_1 = I(X_R; Y_1|X_1) - \beta, \quad (74)$$

so that $\alpha < \beta \leq I(X_R; Y_1|X_1)$. We let $R_0 = R'_0 + \alpha$, $R_1 = R'_1$ and $R_2 = R'_2 - \alpha$. We obtain the following bounds on $(R_0, R_1, R_2)$ based on the bounds on $(R'_0, R'_1, R'_2)$:

$$\begin{aligned} R_0 + R_1 &= R'_0 + R'_1 + \alpha = I(X_R; Y_1|X_1) - \beta + \alpha < I(X_R; Y_1|X_1), \\ R_2 &= R'_2 - \alpha = I(X_D; Y_2|X_1), \\ R_0 + R_2 &= R'_0 + R'_2 < I(X_1, X_D; Y_2). \end{aligned} \quad (75)$$

Hence we have $(R_0, R_1, R_2) \in \mathcal{R}'_{or}$ and, according to Lemma 1, the rate tuple $(R'_0, R'_1, R'_2) = (R_0 - \alpha, R_1, R_2 + \alpha)$ is also achievable. This concludes our proof for the achievability of $\mathcal{C}_{or}$.

# E  Proof of Converse for Theorem 11

Consider a code with length $n$ and average block error probability $P_e$. The probability distribution on the joint ensemble space $W_0 \times W_1 \times \mathcal{X}_a^n \times \mathcal{X}_b^n \times \mathcal{X}_{1a}^n \times \mathcal{X}_{1b}^n \times \mathcal{Y}_{1a}^n \times \mathcal{Y}_{1b}^n \times \mathcal{Y}_{2a}^n \times \mathcal{Y}_{2b}^n$ is given by

$$\begin{aligned} & p(w_0, w_1, x_a^n, x_b^n, x_{1a}^n, x_{1b}^n, y_{1a}^n, y_{1b}^n, y_{2a}^n, y_{2b}^n) \\ & = p(w_0)p(w_1)p(x_a^n, x_b^n|w_0, w_1) \\ & \quad \cdot \prod_{i=1}^n \left[ f_{ai}(x_{1ai}|y_{1a}^{i-1}, y_{1b}^{i-1}) f_{bi}(x_{1bi}|y_{1a}^{i-1}, y_{1b}^{i-1}) p(y_{1ai}, y_{2ai}|x_{ai}, x_{1ai}) p(y_{1bi}, y_{2bi}|x_{bi}, x_{1bi}) \right]. \end{aligned} \quad (76)$$

Note that the channels $p(y_{1ai}, y_{2ai}|x_{ai}, x_{1ai})$ and $p(y_{1bi}, y_{2bi}|x_{bi}, x_{1bi})$ satisfy the degradedness conditions in (33). For convenience, we repeat these conditions here:

$$\begin{aligned} p(y_{1a}, y_{2a}|x_a, x_{1a}) &= p(y_{1a}|x_a, x_{1a})p(y_{2a}|y_{1a}, x_{1a}) \\ p(y_{1b}, y_{2b}|x_b, x_{1b}) &= p(y_{2b}|x_b, x_{1b})p(y_{1b}|y_{2b}, x_{1b}). \end{aligned} \quad (77)$$

We define the following auxiliary random variables

$$T_i := (W_0, Y_{1a}^{i-1}, Y_{1b}^{i-1}, Y_{2a}^{i-1}, Y_{2b}^{i-1}), \qquad \text{for } i = 1, 2, \ldots, n. \quad (78)$$

Note that $T_i$ satisfies the following Markov chain conditions:

$$T_i \to (X_{ai}, X_{1ai}, X_{bi}, X_{1bi}) \to (Y_{1ai}, Y_{2ai}, Y_{1bi}, Y_{2bi}), \qquad \text{for } i = 1, 2, \ldots, n. \quad (79)$$

We first bound the common rate $R_0$:

$$\begin{aligned} nR_0 &= H(W_0) = I(W_0; Y_{2a}^n, Y_{2b}^n) + H(W_0|Y_{2a}^n, Y_{2b}^n) \\ &\overset{(a)}{\leq} I(W_0; Y_{2a}^n, Y_{2b}^n) + n\delta_2 \\ &= \sum_{i=1}^n I(W_0; Y_{2ai}, Y_{2bi}|Y_{2a}^{i-1}, Y_{2b}^{i-1}) + n\delta_2 \\ &= \sum_{i=1}^n I(W_0; Y_{2ai}|Y_{2a}^{i-1}, Y_{2b}^{i-1}) + I(W_0; Y_{2bi}|Y_{2a}^i, Y_{2b}^{i-1}) + n\delta_2 \\ &\overset{(b)}{\leq} \sum_{i=1}^n H(Y_{2ai}) - H(Y_{2ai}|W_0, Y_{1a}^{i-1}, Y_{2a}^{i-1}, Y_{1b}^{i-1}, Y_{2b}^{i-1}, X_{1ai}) \\ &\qquad + H(Y_{2bi}) - H(Y_{2bi}|W_0, Y_{2a}^i, Y_{2b}^{i-1}, X_{bi}, X_{1bi}) + n\delta_2 \\ &\overset{(c)}{\leq} \sum_{i=1}^n I(T_i, X_{1ai}; Y_{2ai}) + H(Y_{2bi}) - H(Y_{2bi}|X_{bi}, X_{1bi}) + n\delta_2 \\ &= \sum_{i=1}^n I(T_i, X_{1ai}, Y_{2ai}) + I(X_{bi}, X_{1bi}; Y_{2bi}) + n\delta_2 \end{aligned} \quad (80)$$

where $(a)$ follows from Fano's inequality, $(b)$ follows because conditioning does not increase entropy, $(c)$ follows from the definition of $T_i$ in (78) and from the Markov chain conditions $(W_0, Y_{2a}^i, Y_{2b}^{i-1}) \to (X_{bi}, X_{1bi}) \to Y_{2bi}$.

We now bound the sum rate $R_0 + R_1$:

$$\begin{aligned}
nR_0 &+ nR_1 \\
&= H(W_0, W_1) = I(W_0, W_1; Y_{1a}^n, Y_{1b}^n) + H(W_0, W_1 | Y_{1a}^n, Y_{1b}^n) \\
&\leq I(W_0, W_1; Y_{1a}^n, Y_{1b}^n) + n\delta_1 \\
&= \sum_{i=1}^n I(W_0, W_1; Y_{1ai}, Y_{1bi} | Y_{1a}^{i-1}, Y_{1b}^{i-1}) + n\delta_1 \\
&\stackrel{(a)}{=} \sum_{i=1}^n I(W_0, W_1; Y_{1ai} | Y_{1a}^{i-1}, Y_{1b}^{i-1}, X_{1ai}) + I(W_0, W_1; Y_{1bi} | Y_{1a}^i, Y_{1b}^{i-1}, X_{1bi}) + n\delta_1 \\
&\stackrel{(b)}{\leq} \sum_{i=1}^n H(Y_{1ai}|X_{1ai}) - H(Y_{1ai}|W_0, W_1, Y_{1a}^{i-1}, Y_{1b}^{i-1}, X_{1ai}, X_{ai}) \\
&\quad + H(Y_{1bi}|X_{1bi}) - H(Y_{1bi}|W_0, W_1, Y_{1a}^i, Y_{1b}^{i-1}, X_{1bi}, X_{bi}) + n\delta_1 \\
&\stackrel{(c)}{=} \sum_{i=1}^n H(Y_{1ai}|X_{1ai}) - H(Y_{1ai}|X_{1ai}, X_{ai}) + H(Y_{1bi}|X_{1bi}) - H(Y_{1bi}|X_{1bi}, X_{bi}) + n\delta_1 \\
&= \sum_{i=1}^n I(X_{ai}; Y_{1ai}|X_{1ai}) + I(X_{bi}; Y_{1bi}|X_{1bi}) + n\delta_1.
\end{aligned} \quad (81)$$

where $(a)$ follows from the chain rule and because $(X_{1ai}, X_{1bi})$ is a function of $(Y_{1a}^{i-1}, Y_{1b}^{i-1})$, $(b)$ follows because conditioning does not increase entropy, and $(c)$ follows from the following Markov chain conditions:

$$\begin{aligned}
(W_0, W_1, Y_{1a}^{i-1}, Y_{1b}^{i-1}) &\to (X_{ai}, X_{1ai}) \to Y_{1ai}, \\
(W_0, W_1, Y_{1a}^i, Y_{1b}^{i-1}) &\to (X_{1bi}, X_{bi}) \to Y_{1bi}.
\end{aligned} \quad (82)$$

We can also derive the following upper bound on the sum rate $R_0 + R_1$:

$$\begin{aligned}
nR_0 &+ nR_1 \\
&= H(W_1) + H(W_0) \\
&\leq I(W_1; Y_{1a}^n, Y_{1b}^n) + I(W_0; Y_{2a}^n, Y_{2b}^n) + n\delta_1 + n\delta_2 \\
&\leq I(W_1; Y_{1a}^n, Y_{2a}^n, Y_{1b}^n, Y_{2b}^n, W_0) + I(W_0; Y_{2a}^n, Y_{2b}^n) + n\delta_1 + n\delta_2 \\
&= \sum_{i=1}^n I(W_1; Y_{1ai}, Y_{2ai}, Y_{1bi}, Y_{2bi} | W_0, Y_{1a}^{i-1}, Y_{2a}^{i-1}, Y_{1b}^{i-1}, Y_{2b}^{i-1}) \\
&\quad + I(W_0; Y_{2ai}, Y_{2bi} | Y_{2a}^{i-1}, Y_{2b}^{i-1}) + n\delta_1 + n\delta_2 \\
&\stackrel{(a)}{\leq} \sum_{i=1}^n I(W_1; Y_{1ai}|T_i, X_{1ai}) + I(W_1; Y_{2ai}|T_i, Y_{1ai}, X_{1ai}) \\
&\quad + I(W_1; Y_{2bi}|T_i, Y_{1ai}, Y_{2ai}, X_{1bi}) + I(W_1; Y_{1bi}|T_i, Y_{1ai}, Y_{2ai}, Y_{2bi}, X_{1bi}) \\
&\quad + I(W_0; Y_{2ai}|Y_{2a}^{i-1}, Y_{2b}^{i-1}) + I(W_0; Y_{2bi}|Y_{2a}^i, Y_{2b}^{i-1}) + n\delta_1 + n\delta_2 \\
&\stackrel{(b)}{\leq} \sum_{i=1}^n I(W_1; Y_{1ai}|T_i, X_{1ai}) + I(W_1; Y_{2bi}|T_i, Y_{1ai}, Y_{2ai}, X_{1bi}) \\
&\quad + I(W_0; Y_{2ai}|Y_{2a}^{i-1}, Y_{2b}^{i-1}) + I(W_0; Y_{2bi}|Y_{2a}^i, Y_{2b}^{i-1}) + n\delta_1 + n\delta_2
\end{aligned} \quad (83)$$

where $(a)$ follows from the chain rule and because $(X_{1ai}, X_{1bi})$ is a deterministic function of $(Y_{1a}^{i-1}, Y_{1b}^{i-1})$, and $(b)$ follows because the second and fourth terms in step $(b)$ are zero due to the degradedness conditions given in (77). The first term in the sum in (83) can be bounded as

$$\begin{aligned}
I(W_1; Y_{1ai}|T_i, X_{1ai}) \\
\leq H(Y_{1ai}|T_i, X_{1ai}) - H(Y_{1ai}|W_1, T_i, X_{1ai}, X_{ai}) \\
\stackrel{(a)}{\leq} H(Y_{1ai}|T_i, X_{1ai}) - H(Y_{1ai}|T_i, X_{1ai}, X_{ai}) \\
\leq I(X_{ai}; Y_{1ai}|T_i, X_{1ai})
\end{aligned} \quad (84)$$

where $(a)$ follows from the Markov chain condition $(W_1, T_i) \to (X_{ai}, X_{1ai}) \to Y_{1ai}$. The third term in the sum in (83) can be bounded as

$$\begin{aligned}
I(W_0; Y_{2ai}|Y_{2a}^{i-1}, Y_{2b}^{i-1}) \\
\leq I(W_0, Y_{1a}^{i-1}, Y_{2a}^{i-1}, Y_{1b}^{i-1}, Y_{2b}^{i-1}, X_{1ai}; Y_{2ai}) \\
= I(T_i, X_{1ai}; Y_{2ai})
\end{aligned} \quad (85)$$

The sum of the second and fourth terms in the sum in (83) can be bounded as

$$\begin{aligned}
I(W_1; Y_{2bi}|T_i, Y_{1ai}, Y_{2ai}, X_{1bi}) + I(W_0; Y_{2bi}|Y_{2a}^i, Y_{2b}^{i-1}) \\
\leq I(W_1; Y_{2bi}|W_0, Y_{1a}^i, Y_{2a}^i, Y_{1b}^{i-1}, Y_{2b}^{i-1}, X_{1bi}) \\
+ I(W_0, Y_{1a}^i, Y_{1b}^{i-1}, X_{1bi}; Y_{2bi}|Y_{2a}^i, Y_{2b}^{i-1}) \\
= I(W_0, W_1, Y_{1a}^i, Y_{1b}^{i-1}, X_{1bi}; Y_{2bi}|Y_{2a}^i, Y_{2b}^{i-1}) \\
\leq H(Y_{2bi}) - H(Y_{2bi}|W_0, W_1, Y_{1a}^i, Y_{1b}^{i-1}, Y_{2a}^i, Y_{2b}^{i-1}, X_{1bi}, X_{bi}) \\
\stackrel{(a)}{=} H(Y_{2bi}) - H(Y_{2bi}|X_{bi}, X_{1bi}) \\
= I(X_{bi}, X_{1bi}; Y_{2bi})
\end{aligned} \quad (86)$$

where $(a)$ follows from the Markov chain condition $(W_0, W_1, Y_{1a}^i, Y_{1b}^{i-1}, Y_{2a}^i, Y_{2b}^{i-1}) \to (X_{bi}, X_{1bi}) \to Y_{2bi}$. Substituting (84), (85), and (86) into (83), we obtain

$$\begin{aligned}
nR_0 + nR_1 \\
\leq \sum_{i=1}^n I(X_{ai}; Y_{1ai}|T_i, X_{1ai}) + I(T_i, X_{1ai}; Y_{2ai}) + I(X_{bi}, X_{1bi}; Y_{2bi}) + n\delta_1 + n\delta_2
\end{aligned} \quad (87)$$

Finally, we note that all the terms in the bounds (80), (81) and (87) are determined either by the distribution $p(t_i, x_{ai}, x_{1ai}, y_{1ai}, y_{2ai})$ or by the distribution $p(x_{bi}, x_{1bi}, y_{1bi}, y_{2bi})$. Hence there is no loss of generality to consider only joint distributions of the form $p(t_i, x_{ai}, x_{1ai})p(x_{bi}, x_{1bi})$ for $(T_i, X_{ai}, X_{1ai}, X_{bi}, X_{1bi})$.

The converse for Theorem 11 now follows by using standard arguments (see the end of Appendix B and e.g. [24, p. 402]).

# Acknowledgement

The authors would like to thank Professor Venugopal Veeravalli for supporting Yingbin Liang's visit at Bell Labs in January 2005 where much of this work was performed.